\documentclass[12pt,a4paper,superscriptaddress,preprint]{revtex4}
\usepackage[dvips]{graphicx}
\usepackage{amssymb,amsfonts,amsmath}
\usepackage{color}
\usepackage{bm}
\usepackage{color}

\newcommand{\green}[1]{}

\begin{document}

\newcommand{\EQ}{Eq.~}
\newcommand{\EQS}{Eqs.~}
\newcommand{\FIG}{Fig.~}
\newcommand{\FIGS}{Figs.~}
\newcommand{\TAB}{Tab.~}
\newcommand{\TABS}{Tabs.~}
\newcommand{\SEC}{Sec.~}
\newcommand{\SECS}{Secs.~}

\title{Dynamics-based centrality for general directed networks}
\author{Naoki Masuda}
\affiliation{Department of Mathematical Informatics,
The University of Tokyo,
7-3-1 Hongo, Bunkyo, Tokyo 113-8656, Japan}
\affiliation{PRESTO, Japan Science and Technology Agency,
4-1-8 Honcho, Kawaguchi, Saitama 332-0012, Japan}
\author{Hiroshi Kori}
\affiliation{Division of Advanced Sciences, Ochadai Academic Production,
Ochanomizu University,
2-1-1, Ohtsuka, Bunkyo-ku, Tokyo 112-8610, Japan}
\affiliation{PRESTO, Japan Science and Technology Agency,
4-1-8 Honcho, Kawaguchi, Saitama 332-0012, Japan}

\begin{abstract}
Determining the relative importance of nodes in directed networks is important in, for example, ranking websites, publications, and sports teams, and for understanding signal flows in systems biology. A prevailing centrality measure in this respect is the PageRank. In this work, we focus on another class of centrality derived from the Laplacian of the network. We extend the Laplacian-based centrality, which has mainly been applied to strongly connected networks, to the case of general directed networks such that we can quantitatively compare arbitrary nodes. Toward this end, we adopt the idea used in the PageRank to introduce global connectivity between all the pairs of nodes with a certain strength. Numerical simulations are carried out on some networks. We also offer interpretations of the Laplacian-based centrality for general directed networks in terms of various dynamical and structural properties of networks. Importantly, the Laplacian-based centrality defined as the stationary density of the continuous-time random walk with random jumps is shown to be equivalent to the absorption probability of the random walk with sinks at each node but without random jumps. Similarly, the proposed centrality represents the importance of nodes in dynamics on the original network supplied with sinks but not with random jumps.
\end{abstract}

\maketitle

\newpage

\section{Introduction}\label{sec:introduction}

A network is a set of nodes and a set of links that connect pairs of
nodes (see \cite{albert02rmp,Boccaletti06,Newman10book} for reviews).
In applications including information science, sociology, and
biology, it is often necessary to determine important nodes in a network.
Various definitions of the 
importance of nodes, or centrality measures,
have been proposed since the first 
classical studies on social network
analysis in the 1950s
\cite{Katz53,Boccaletti06,Wasserman94-Brandes05LNCSbook}.

It is often more suitable to consider links 
to be directed, where the direction of
link represents relationships such as 
the control of one node over another, unidirectional flow, and
citation.
Many centrality measures including degree centrality,
betweenness centrality, and eigenvector centrality can be
adopted to the case of directed networks.
Nevertheless, the most
popular centrality for directed networks appears to be PageRank,
which takes nontrivial values only in directed networks.
It was originally developed for ranking websites
\cite{Brin98-Langville06book}. In other words, the PageRank of a node is
large when the node receives many links from important nodes that do not
have too many outgoing links.

In the present study, we focus on another important class of
centrality for directed networks, i.e., those derived from the
Laplacian of the network. This class of centrality has a long history
\cite{Daniels69Biom-Berman80SIAM,Moon70SiamRev,Biggs97BullLondMathSoc,Agaev00,Chebotarev05arxiv,Borm02}
and is mathematically close to the PageRank (see
\SEC\ref{sec:our def}).  Furthermore, for strongly
connected networks, i.e., networks in which there exists a path of
directed links between an arbitrary ordered pair of nodes, the
Laplacian-based centrality value of a node, which we also call the
influence of a node, represents
its importance in
various dynamics on networks \cite{mkk09njp,mkk09pre,Klemm10arxiv}.

The Laplacian-based centrality measure has mostly been analyzed for
strongly connected networks
\cite{Daniels69Biom-Berman80SIAM,Moon70SiamRev,Biggs97BullLondMathSoc,mkk09njp,mkk09pre,Klemm10arxiv}.
However, real directed networks may not be strongly connected. This is
typically the case when the network is sparse (i.e., number of
links is relatively small) or of small size.  Although the
Laplacian-based centrality in the original form is applicable when all
the nodes are reached along directed paths from a certain specified
node, such a network is not generic.  The Laplacian-based centrality
has been generalized to the case of general directed networks
\cite{Agaev00,Borm02,Chebotarev05arxiv}.  In the generalized version,
nodes in an uppermost component have positive centrality values,
whereas nodes in a downstream component have zero centrality values
(see \SECS\ref{sec:def for strongly connected networks} and
\ref{sec:def for general networks} for definitions of uppermost and
downstream components).  However, we may want to compare the
importance of nodes in downstream components. We may also wish to
compare a node 1 in an uppermost component and a node 2 in a
downstream component that is not under the control of node 1.

In this paper, we extend the
Laplacian-based centrality measure (i.e., influence) to the case of
general directed networks. Networks do not have to be strongly connected
and can be composed of disconnected components.
The extended centrality measure,
called as the influence or extended influence
without ambiguity,
is a one-parameter family of the centrality measure with parameter
$q$ such that
the previous definition \cite{Agaev00,Borm02,Chebotarev05arxiv} is
recovered in the limit
$q\to 0$. The extended influence is a relative of the PageRank;
the influence and the PageRank correspond to
continuous-time and discrete-time simple random walks, respectively.
The present paper is organized as follows.
In \SECS\ref{sec:def for strongly connected networks}
and \ref{sec:def for general networks},
we review previous works on the influence for strongly connected
and general directed networks, respectively.
In \SEC\ref{sec:interpretation}, we present new interpretations of
the centrality 
measure introduced in \SEC\ref{sec:def for general networks}.
In \SEC\ref{sec:our def}, we extend the concept of the influence by
borrowing the idea used in the PageRank to introduce some global
connectivity to the original network.
We also show that the proposed influence can be interpreted
as the dynamical properties of nodes on the original network 
without additional global connectivity.
In \SECS\ref{sec:toy examples} and \ref{sec:large-N examples},
we apply the influence to toy examples and relatively
large networks, respectively.
In \SEC\ref{sec:conclusions}, we summarize and discuss our results, with
an emphasis on the comparison of the influence and the PageRank.

\section{Influence for networks with single zero Laplacian
  eigenvalue}\label{sec:def for strongly connected networks}

Consider a directed and weighted network having $N$ nodes.  The weight
of the link from node $i$ to node $j$ is denoted by $w_{ij}$ and
assumed to be nonnegative.  $w_{ij}>0$ represents the strength with
which node $i$ governs node $j$.  $w_{ij}$ and $w_{ji}$ are
generally different from each other.

The Laplacian-based centrality measure, called the
influence of node $i$ and denoted as $v_i$,
is defined as the solution of the following set of $N$ linear equations:
\begin{equation}
v_i = \frac{\sum_{j=1}^N w_{ij}v_j}
{\sum_{j^{\prime}=1}^N w_{j^{\prime} i}},\quad (1\le i\le N).
\label{eq:influence}
\end{equation}
The normalization is given by $\sum_{i=1}^N v_i=1$.
We can rewrite \EQ\eqref{eq:influence} as
\begin{equation}
(v_1\; \cdots\; v_N) L = 0,
\label{eq:influence with L}
\end{equation}
where 
$L=(L_{ij})$ is the asymmetric Laplacian defined by 
\begin{equation}
L_{ij}=\delta_{ij}\sum_{j^{\prime}\neq i}w_{j^{\prime}i}
-(1-\delta_{ij})w_{ji}.
\end{equation}
$v_i$ represents the importance of nodes in various dynamics on
networks, such as the voter model, a random walk, DeGroot's model
of consensus formation, and the response of synchronized networks
\cite{mkk09njp}.

If a network is strongly connected, that is, if any node $j$ can be
reached from an arbitrary node $i$ along directed links, the
Perron-Frobenius theorem guarantees that $(v_1 \; \cdots\; v_N)$ is
unique and $v_i>0$ ($1\le i\le N$).  In particular, for undirected
networks, which are strongly connected as long as they are connected,
we have $v_i=1/N$.  Therefore, the influence is a centrality measure
that is relevant only in directed networks.

\begin{figure}
\begin{center}
\includegraphics[width=6cm]{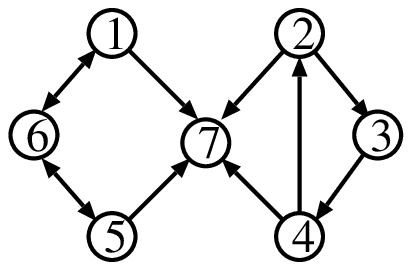}
\caption{A network with two root nodes.}
\label{fig:example Gr}
\end{center}
\end{figure}

\begin{figure}
\begin{center}
\includegraphics[width=8cm]{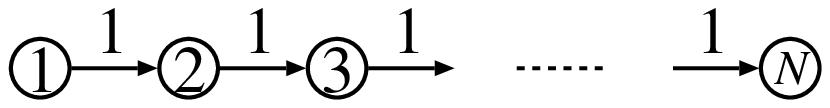}
\caption{Directed chain with $N$ nodes.}
\label{fig:directed chain}
\end{center}
\end{figure}

To discuss the uniqueness of the zero eigenvector $(v_1\; \cdots \;
v_N)$ of $L$, we use the concept of the root node \cite{Agaev00}.
Consider the set $G_{\rm r}$ of $m$ nodes in a given network ($1\le
m\le N$).  We define $G_{\rm r}$ to be a set of root nodes if an
arbitrary node can be reached along directed links from a node
included in $G_{\rm r}$ and $G_{\rm r}$ is minimal. In the example
shown in \FIG\ref{fig:example Gr}, $\{1, 2\}$ qualifies as
$G_{\rm r}$. $\{1, 3\}$ is another example of $G_{\rm r}$. $\{1,
4, 5\}$ does not qualify because it is not minimal.  The
minimality indicates that some nodes cannot be reached from $G_{\rm
r}^{\prime}$, where $G_{\rm r}^{\prime}$ is the set of nodes with
$m-1$ nodes defined by removing an arbitrary node from $G_{\rm r}$.
For strongly connected networks ($m=1$), $G_{\rm r}$ can be
a set of any single node. As this exercise suggests, $G_{\rm r}$ for
a given network is generally not unique. However, $m$ is uniquely
determined from a network \cite{Agaev00}. The directed chain
shown in \FIG\ref{fig:directed chain} is a network that is not strongly
connected with $m=1$. In \FIG\ref{fig:directed chain}, we obtain $v_1=1$ for
the unique root node 1 and $v_i=0$ ($2\le i\le N$).

The multiplicity of the zero eigenvalue of $L$,
also called the geometric multiplicity of the eigenvalue
\cite{Golub96book,Hornbook}, is equal to $m$
\cite{Abelson64chap,ermentrout92,Agaev00}.
Therefore, the influence given by \EQ\eqref{eq:influence with L}
is well defined only for networks with $m=1$, and most
previous papers that treat \EQ\eqref{eq:influence with L}
concentrate on strongly connected networks
\cite{Daniels69Biom-Berman80SIAM,Moon70SiamRev,Biggs97BullLondMathSoc,MasudaOhtsuki09njp,mkk09njp,mkk09pre,Klemm10arxiv}.
In this case, $(v_1\; \cdots\; v_N)$
can be readily calculated
by the power iteration or the enumeration of the directed
spanning tree \cite{mkk09njp,mkk09pre}.

\section{Case of multiple zero Laplacian eigenvalues}\label{sec:def
  for general networks}

In this section, we treat networks with multiple zero Laplacian
eigenvalues.  Such a network is not strongly connected.  The
influence explained in \SEC\ref{sec:def for strongly connected networks}
was extended to accommodate this case
by Agaev--Chebotarev \cite{Agaev00,Chebotarev05arxiv}
and Borm et al. \cite{Borm02}.
We develop a new centrality measure
in \SEC\ref{sec:our def} by generalizing their definitions.
In this section, we explain their
centrality measure and
examine its properties.

Consider a continuous-time simple random walk on the network generated
by reversing the direction of all the links of the original
network. We select each node $i$ ($1\le i\le N$)
as the initial
location of the random walker with probability $1/N$. For
directed networks that are not necessarily strongly connected,
Agaev--Chebotarev \cite{Agaev00,Chebotarev05arxiv} and
Borm et al. \cite{Borm02} defined a centrality measure, which we call the
influence and denote by
$v_i$ without ambiguity, as the
long-term probability that the walker visits node $i$.
For a strongly connected network, $v_i$ is equal to the stationary
density of the random walk and coincides with $v_i$ defined by
\EQ\eqref{eq:influence with L}
\cite{mkk09njp,MasudaOhtsuki09njp}.  For a network with a single root
node $i_0$, node $i_0$ is the unique absorbing boundary, and any
random walker is eventually trapped at node $i_0$. Therefore,
$v_{i_0}=1$ and $v_i=0$ ($i\neq i_0$), which is again consistent
with \EQ\eqref{eq:influence with L} \cite{mkk09njp}.

Because the generator of the continuous-time random walk is equal to
$-L$, we obtain
\begin{equation}
(v_1\; \cdots \; v_N)=
\lim_{t\to\infty}\frac{1}{N}(1\; \cdots\; 1)\exp(-Lt).
\label{eq:derivation of extended v_i}
\end{equation}

The spectral decomposition of $L$ yields
\begin{equation}
e^{-Lt}= \sum_{m^{\prime}=1}^m\bm u^{(0)}_{m^{\prime}} 
\bm v^{(0)}_{m^{\prime}} + 
\left(\bm u^{(\lambda_2)}_1 \bm v^{(\lambda_2)}_1 + \ldots
\right)e^{-\lambda_2 t} + \ldots,
\label{eq:spectral decomposition of L}
\end{equation}
where $\bm v^{(0)}_{m^{\prime}}$ and $\bm u^{(0)}_{m^{\prime}}$ 
($1\le m^{\prime}\le m$) are
the zero left and right eigenvectors of $L$, respectively.
Without loss of generality, we assume that
$\bm v^{(0)}_{m^{\prime}}$ 
and $\bm u^{(0)}_{m^{\prime}}$ are normalized and orthogonalized
such that
\begin{equation}
\sum_{j=1}^N v^{(0)}_{m^{\prime},j}=1,\quad (1\le m^{\prime}\le m),
\label{eq:normalize}
\end{equation}
and 
\begin{equation}
\bm v^{(0)}_{m_1} \bm u^{(0)}_{m_2}=\delta_{m_1 m_2}.
\label{eq:orthogonalize}
\end{equation}
$\lambda_2$ is the spectral gap (i.e., smallest positive
eigenvalue) of $L$. $\bm u^{(\lambda_2)}_1$ and $\bm
v^{(\lambda_2)}_1$ are a pair of left and right eigenvalues of $L$
corresponding to $\lambda_2$, where 
$\bm v^{(\lambda_2)}_1 \bm u^{(\lambda_2)}_1\neq 0$.
Other modes that decay at least as fast as 
$e^{-\lambda_2 t}$ with $t$ are omitted in \EQ\eqref{eq:spectral decomposition of L}.
Note that \EQ\eqref{eq:spectral decomposition of L} 
is also valid when $L$ is not 
diagonalizable and has a nondiagonal Jordan normal form.
The combination of \EQS\eqref{eq:derivation of extended v_i}
and \eqref{eq:spectral decomposition of L}
leads to
\begin{equation}
(v_1\; \cdots\; v_N) = \frac{1}{N}(1\; \cdots\; 1)
\sum_{m^{\prime}=1}^m\bm u^{(0)}_{m^{\prime}} \bm v^{(0)}_{m^{\prime}}.
\label{eq:def of extended v_i}
\end{equation}

When $m=1$, by substituting $\bm u^{(0)}_1=
(1\; \cdots\; 1)^{\top}$ ($\top$ denotes the transpose)
into \EQ\eqref{eq:def of extended v_i}, we obtain
$(v_1\; \cdots\; v_N)=\bm v^{(0)}_1$. Therefore,
\EQ\eqref{eq:def of extended v_i} extends \EQ\eqref{eq:influence with L}.
We note that \EQ\eqref{eq:def of extended v_i} is also applicable to
disconnected networks for which $m\ge 2$.

To gain insights into \EQ\eqref{eq:def of extended v_i},
we consider the decomposition of 
directed networks into strongly connected
components (SCCs).
We define the
uppermost SCC as an SCC that is not downstream to any other SCC
along directed links. The number of uppermost SCCs in a given network is 
equal to $m$ \cite{Agaev00}.
The choice of $G_{\rm r}$, the set of root nodes,
is unique up to the arbitrariness of the choice of
a node in each uppermost SCC. This is consistent with the fact that
the set of any single node is qualified as $G_{\rm r}$ in a strongly
connected network.

In general, we can permute the indices of the nodes
such that $L$
is the irreducible normal form \cite{Hornbook,Moon70SiamRev} given by
\begin{equation}
L=\left(\begin{array}{ccccc}
L_{11}&0     &0     &\cdots&0\\
L_{21}&L_{22}&0     &\cdots&0\\
L_{31}&L_{32}&L_{33}&\cdots&\vdots\\
\vdots&\vdots&\vdots&\ddots&0\\
L_{b1}&L_{b2}&L_{b3}&\cdots&L_{bb}
\end{array}\right).
\label{eq:L irreducible normal form}
\end{equation}
The diagonal block $L_{b^{\prime}b^{\prime}}$ ($1\le b^{\prime}\le b$)
corresponds to the $b^{\prime}$th SCC.
We denote the number of nodes in the $b^{\prime}$th SCC by
$N_{b^{\prime}}$. Then, 
$L_{b^{\prime}b^{\prime\prime}}$ is an $N_{b^{\prime}}
\times N_{b^{\prime\prime}}$ matrix and
$\sum_{b^{\prime}=1}^b N_{b^{\prime}}=N$.
The lower triangular nature of \EQ\eqref{eq:L irreducible normal form}
implies that the SCCs are ordered in 
\EQ\eqref{eq:L irreducible normal form} such that
links may exist from a node in the $b^{\prime}$th SCC to
the $b^{\prime\prime}$th SCC only when $b^{\prime}\le b^{\prime\prime}$.

Because $m$ out of $b$
SCCs do not receive links from other SCCs, the uppermost SCCs occupy
the first $m$ rows of 
blocks in \EQ\eqref{eq:L irreducible normal form}, and
we obtain
\begin{equation}
L_{m_1 m_2}=0,\quad (m_1>m_2, 1\le m_1\le m).
\label{eq:cnd for Frobenius normal form}
\end{equation}
Equation~\eqref{eq:L irreducible normal form} constrained by
\EQ\eqref{eq:cnd for Frobenius normal form} is
called the Frobenius normal form \cite[p.~38]{Bapat97book}.
In addition, $L_{m^{\prime}m^{\prime}}$ ($1\le m^{\prime}
\le m$) is the Laplacian matrix
of the $m^{\prime}$th SCC, which has a single zero eigenvalue. The
eigenequations for this submatrix are represented by
\begin{equation}
L_{m^{\prime}m^{\prime}}\left(\begin{array}{c}
1\\ \vdots\\ 1
\end{array}\right)=0,
\quad
(v_{m^{\prime},1}\; \cdots\; v_{m^{\prime},N_{m^{\prime}}}) 
L_{m^{\prime}m^{\prime}}=0.
\end{equation}

It is easy \cite{Agaev00}
to verify that the $m$ left zero eigenvectors of $L$ are
given by
\begin{equation}
\bm v^{(0)}_{m^{\prime}} = 
(\underbrace{0\; \cdots\; 0}_{\begin{subarray}{c}
N_1+\ldots + N_{m^{\prime}-1}\\\text{ zeros}\end{subarray}}\; 
v_{m^{\prime},1}\; v_{m^{\prime},2}\; \cdots\;
v_{m^{\prime},N_{m^{\prime}}}\; 
\underbrace{0\; \cdots\; 0}_{\begin{subarray}{c}
N_{m^{\prime}+1}+\ldots +N_b\\ \text{ zeros}\end{subarray}}),
\quad \sum_{j=1}^{N_{m^{\prime}}}
v_{m^{\prime},j}=1, \quad (1\le m^{\prime}\le m).
\label{eq:elements of v0}
\end{equation}
To satisfy the normalization condition
$\bm v^{(0)}_{m_1} \bm u^{(0)}_{m_2}=\delta_{m_1,m_2}$
and the first $\sum_{m^{\prime}=1}^m N_{m^{\prime}}$ 
rows of $L\bm u^{(0)}_{m^{\prime}} = 0$,
we should take
\begin{equation}
\bm u^{(0)}_{m^{\prime}} = 
(\underbrace{0\; \cdots\; 0}_{\begin{subarray}{c}
N_1+\ldots + N_{m^{\prime}-1}\\\text{ zeros}\end{subarray}}\; 
\underbrace{1\; \cdots\; 1}_{N_{m^{\prime}} \text{ ones}}
\underbrace{0\; \cdots\; 0}_{\begin{subarray}{c}
N_{m^{\prime}+1}+\ldots +N_m\\ \text{zeros}\end{subarray}}
\overline{\bm u}^{m+1 \top}_{m^{\prime}}\; \cdots\;
\overline{\bm u}^{b \top}_{m^{\prime}})^{\top},
\label{eq:u0 for m>=2}
\end{equation}
where
$\overline{\bm u}^{b^{\prime}}_{m^{\prime}}$ 
($m+1\le b^{\prime} \le b$) is the
$N_{b^{\prime}}$-dimensional column vector determined by
\begin{equation}
\left(\begin{array}{c}
L_{m+1,m^{\prime}}\\ \vdots\\ L_{b,m^{\prime}}\end{array}\right)
\left(\begin{array}{c}
1\\ \vdots\\ 1\end{array}\right) +
\left(\begin{array}{ccccc}
L_{m+1,m+1}&0     &\cdots&0\\
L_{m+2,m+1}&L_{m+2,m+2}&\cdots&0\\
\vdots&\vdots&\ddots&0\\
L_{b,m+1}&L_{b,m+2}&\cdots&L_{bb}
\end{array}\right)
\left(\begin{array}{c}
\overline{\bm u}^{m+1}_{m^{\prime}}\\ \vdots\\ 
\overline{\bm u}^b_{m^{\prime}}
\end{array}\right)=0.
\label{eq:u rest}
\end{equation}

To show that \EQ\eqref{eq:u rest} has a unique nonnegative solution,
we decompose the diagonal block $L_{m^{\prime}m^{\prime}}$ 
($m+1\le m^{\prime}\le N$)
as 
\begin{equation}
L_{m^{\prime}m^{\prime}}=\tilde{L}_{m^{\prime}m^{\prime}}+D_{m^{\prime}},
\label{eq:L diagonal lower}
\end{equation}
where $\tilde{L}_{m^{\prime}m^{\prime}}$ is the Laplacian of the
$m^{\prime}$th SCC and $D_{m^{\prime}}$ is the diagonal matrix whose
$i$th element is equal to the total number of incoming links from SCCs
1, $\ldots$, $m^{\prime}-1$ to the $i$th node in the $m^{\prime}$th
SCC.  Equation~\eqref{eq:L diagonal lower} implies that
$L_{m^{\prime}m^{\prime}}$ is diagonally dominant.  Therefore, by
applying the Jacobi or Gauss-Seidel iteration to the first $N_{m+1}$
rows of \EQ\eqref{eq:u rest}, we can uniquely calculate $
\overline{\bm u}^{m+1}_{m^{\prime}}$ \cite{Golub96book}.
Furthermore, $L_{m^{\prime}m^{\prime}}$ is an M-matrix \cite{BermanPlemmons79book}.
Because all the elements of $L_{m+1,m^{\prime}}(1\; \cdots\;
1)^{\top}$ that appear in the first $N_{m+1}$ rows of \EQ\eqref{eq:u
rest} are not positive, all the elements of 
$\overline{\bm u}^{m+1}_{m^{\prime}}$
are guaranteed to be nonnegative \cite[p.~136]{BermanPlemmons79book}.
By substituting the obtained
$\overline{\bm u}^{m+1}_{m^{\prime}}$ in \EQ\eqref{eq:u rest} and
applying the Jacobi or Gauss-Seidel iteration to the next $N_{m+2}$
rows, we can uniquely determine $\overline{\bm u}^{m+2}_{m^{\prime}}$.
By repeating the same procedure, we can successively determine
$\overline{\bm u}^{(0)}_{m^{\prime}}$, whose elements are unique and
nonnegative.

The projection of $L$ onto the eigenspaces yields
\begin{equation}
I= \left(\sum_{m^{\prime}=1}^m\bm u^{(0)}_{m^{\prime}} \bm v^{(0)}_{m^{\prime}}\right) + 
\bm u^{(\lambda_2)}_1 \bm v^{(\lambda_2)}_1 + \ldots,
\label{eq:L projection}
\end{equation}
where $I$ is the $N\times N$ unit matrix. Note that 
\EQ\eqref{eq:L projection} is valid even if $L$ is not diagonalizable.
By multiplying the $N$-dimensional column
vector $(1\; \cdots\; 1)^{\top}$, a zero right eigenvector of $L$,
from the right to both sides of \EQ\eqref{eq:L projection} and using
\EQS\eqref{eq:normalize} and \eqref{eq:orthogonalize}, we obtain
\begin{equation}
\left(\begin{array}{c}
1\\ \vdots\\ 1 \end{array}\right)=
\sum_{m^{\prime}=1}^m \bm u^{(0)}_{m^{\prime}}.
\label{eq:sum to unity}
\end{equation}
Equation~\eqref{eq:sum to unity} implies
$\sum_{m^{\prime}=1}^m u^{(0)}_{m^{\prime},j}=1$ ($1\le j\le
N$), where $u^{(0)}_{m^{\prime},j}$ is the $j$th element of
$\bm u^{(0)}_{m^{\prime}}$ and represents
the probability that the random walk starting from node $j$
is trapped by the $m^{\prime}$th uppermost SCC.
$u^{(0)}_{m^{\prime},j}$ can be interpreted as the magnitude of
the influence that the $m^{\prime}$th SCC
exerts on node $j$.
Note that $u^{(0)}_{m^{\prime},j}=0$ if  
node $j$ cannot be reached from the $m^{\prime}$th uppermost SCC 
along directed links in the original network.

By substituting \EQ\eqref{eq:elements of v0} in
\EQ\eqref{eq:def of extended v_i}, we obtain
\begin{equation}
v_i = v_{m^{\prime},i} \frac{\sum^N_{j=1} u^{(0)}_{m^{\prime},j}}{N}
\label{eq:def influence general}
\end{equation}
for node $i$ that belongs to the $m^{\prime}$th uppermost SCC.
For these nodes, $v_i>0$ is satisfied.
For nodes that do not belong to an uppermost SCC, 
we obtain $v_i=0$. Equation~\eqref{eq:sum to unity} guarantees
that $\sum_{i=1}^N v_i=1$.
Equation~\eqref{eq:def influence general} generalizes
the definition for strongly connected networks
given by \EQ\eqref{eq:influence}.
We interpret the right-hand side of 
\EQ\eqref{eq:def influence general} to be the multiplication
of the influence of node $i$ within the $m^{\prime}$th SCC (i.e., $v_{m^{\prime},i}$)
and the relative influence of the $m^{\prime}$th SCC in the entire network
(i.e., $\sum^N_{j=1} u^{(0)}_{m^{\prime},j}/N$).

For pedagogical purposes, the calculations of $v_i$ for two
toy networks with $N=4$ and $m=2$ are presented in the Appendix.

\section{Interpretation of influence for networks with 
multiple root nodes}\label{sec:interpretation}

Borm and colleagues defined the Laplacian 
centrality measure on the basis of the continuous-time simple random
walk on networks. In this section, we further motivate this definition
by showing that $v_i$ given by
\EQ\eqref{eq:def influence general}
have other interpretations,
as is the case for $v_i$ formulated for strongly connected
networks \cite{mkk09njp}.

\subsection{Collective responses in the DeGroot model of consensus
formation}

The DeGroot model represents dynamical opinion formation in a
population of interacting individuals
\cite{Degroot74-Friedkin91}. The dynamics of the continuous-time
version of the DeGroot model \cite{olfatisaber07},
also known as Abelson's model \cite{Abelson64chap},
are defined by
\begin{equation}
\dot{\bm x} (t) = -L \bm x (t),
\end{equation}
where $\bm x(t)\equiv \left[x_1(t)\; \cdots \;
x_N(t)\right]^{\top}\in \bm R^N$ 
represents the time-dependent opinion vector.
For networks with $m=1$, including
strongly connected networks, the consensus, i.e.,
synchrony, is asymptotically reached. In this case,
the final synchronized opinion is given by
$\bm v \bm x(0)$. Therefore, 
$v_i$ is equal to the fraction of
the initial opinion at node $i$ reflected in the final
opinion of the entire network \cite{Degroot74-Friedkin91,olfatisaber07,kori09,mkk09njp}.

When $m\ge 2$, synchrony is neutrally but not asymptotically
stable. Therefore,
the consensus of the entire
network is not generally reached from 
a general initial condition. The final opinion vector
is given by
\begin{equation}
\lim_{t\to\infty}\bm x(t)=\lim_{t\to\infty}e^{-Lt}\bm x(0)=
\left(\sum_{m^{\prime}=1}^m\bm u^{(0)}_{m^{\prime}} \bm
v^{(0)}_{m^{\prime}}
\right)\bm x(0).
\label{eq:linear solution 3}
\end{equation}
If we set $x_j(0)=\delta_{ij}$ ($1\le j\le N$) to
introduce a different opinion of unit strength at node $i$
to the initial all-0 consensus state,
the average response of the nodes induced by a different
opinion at node $i$ is equal to
\begin{equation}
\lim_{t\to\infty}\frac{1}{N}\sum_{j=1}^N x_j(t)=
\frac{1}{N}(1\; \cdots\; 1)
\sum_{m^{\prime}=1}^m\bm u^{(0)}_{m^{\prime}} \bm
v^{(0)}_{m^{\prime}}\bm e_i,
\label{eq:LD stationary}
\end{equation}
where $\bm e_i$ is the $N$-dimensional unit column vector such that 
the $i$th element is equal to 1 and the other elements are equal to 0.
Because \EQ\eqref{eq:LD stationary}
coincides with \EQ\eqref{eq:def influence general},
the amount of the initial opinion of
node $i$ reflected in the final opinion of the
entire network is given by
\EQ\eqref{eq:def influence general}.

\subsection{Stationary density of voter model}

The so-called link dynamics is a stochastic interacting particle system on networks
in which each node takes one of the two opinions $A$ and $B$
\cite{Antal06prl-Sood08pre}.  In each time step, one link is randomly
selected from the network with a probability proportional to the
weight of the link. Then, the state of the source node of the link
replaces that of the target node of the link if their states are different.
Note that opinions $A$ and $B$ are equally strong in the dynamics. 
The dynamics halt when $A$ or $B$
takes over the entire network. The fixation
probability of node $i$ is defined as the probability that $B$
takes over the network when the initial configuration is such that 
node $i$ takes $B$ and the other $N-1$ nodes take $A$.
When $m=1$, $v_i$ is equal to
the fixation probability of node $i$
\cite{MasudaOhtsuki09njp,mkk09njp}.

When $m\ge 2$, the fixation of $B$ introduced at
node $i$ never occurs. If node $i$ is
located in a downstream SCC, $B$ eventually vanishes because
$A$ in the uppermost SCCs is permanent and replaces $B$
in the downstream SCCs. If node $i$ is located in an uppermost SCC, this
SCC ends up with being
entirely occupied by $B$ with a positive probability. However,
other uppermost SCCs are permanently occupied by $A$, such that
the consensus is never reached.

In this situation, consider the expected fraction of
$B$ in the network in the stationary state when we start from the
initial configuration with a single $B$ at node $i$.
The probability that $B$
takes over the $m^{\prime}$th uppermost SCC to which node $i$ belongs is equal to
$v_{m^{\prime},i}$. Under the condition that the $m^{\prime}$th
uppermost SCC is entirely occupied by $B$ and the other $m-1$
uppermost SCCs are entirely occupied by $A$, the master equation for
the
probability $p_j^{\rm LD}$ that node $j$ in a downstream SCC is occupied by $B$
is given by
\begin{align}
\frac{d p_j^{\rm LD}}{dt} =& (1-p_j^{\rm LD})
\sum_{j^{\prime}=1}^N w_{j^{\prime} j}
p_{j^{\prime}}^{\rm LD} - 
p_j^{\rm LD} \sum_{j^{\prime}=1}^N w_{j^{\prime} j} 
(1-p_{j^{\prime}}^{\rm LD}),\notag\\
=& \sum_{j^{\prime}=1}^N w_{j^{\prime} j} p_{j^{\prime}}^{\rm LD} -
p_j^{\rm LD} \sum_{j^{\prime}=1}^N w_{j^{\prime} j},
\label{eq:master equation for LD}
\end{align}
where we set $p_{j^{\prime}}^{\rm LD}=1$ when node $j^{\prime}$
belongs to the $m^{\prime}$th uppermost SCC and 
$p_{j^{\prime}}^{\rm LD}=0$ when node $j^{\prime}$
belongs to one of the other uppermost SCCs.

Equation~\eqref{eq:master equation for LD} implies that in the equilibrium,
$(p_1^{\rm LD}\; \cdots\;  p_N^{\rm LD})^{\top}$ is a right zero
eigenvector of $L$ and identical to
$\bm u^{(0)}_{m^{\prime}}$ given by \EQ\eqref{eq:u0 for m>=2}.
Therefore, the stationary fraction of opinion $B$ in the network is
given by \EQ\eqref{eq:def influence general}.

\subsection{Enumeration of spanning trees}

When $m=1$, the matrix-tree theorem implies that
$v_i$ is proportional to
the sum of the weights of all the possible directed spanning trees
rooted at node $i$ \cite{Biggs97BullLondMathSoc,Agaev00,mkk09njp}.
The weight of a spanning tree is defined as the multiplication of all the
weights of the $N-1$ links included in the spanning tree.

The Markov chain tree theorem extends this result to the case $m\ge 2$
\cite{Leighton86IEEE}.  According to this theorem,
$v_{m^{\prime},i}u^{(0)}_{m^{\prime},j}$ for general directed networks
is proportional to the sum of the weights of all the arborescences
such that node $i$ is a root node of the arborescence and the
arborescence passes node $j$.  An arborescence is a subgraph of the
original networks with $N$ nodes such that the indegree of each node
restricted to the arborescence is at most one, it has no cycles, and
it contains the maximal number of links.  The nodes whose indegrees
are zero within the arborescence are called the root nodes of the
arborescence. They form $G_{\rm r}$ such that the concept of the root
node for the arborescence and that for the network Laplacian
\cite{Agaev00} are identical. Therefore, the number of links in an
arborescence is equal to $N-m$, and the arborescence is composed of
$m$ disconnected directed trees each of which emanates from a root
node. Intuitively, $v_{m^{\prime},i}u^{(0)}_{m^{\prime},j}$ represents
the number of different ways in which node $i$ influences node $j$.

The influence of node $i$ defined by
\EQ\eqref{eq:def influence general} is proportional to
the summation of all the arborescences with the modified weight.
The modified weight of an arborescence 
is defined by the multiplication of all the weights of
the $N-m$ links included in the arborescence and the number of nodes
included in the directed tree rooted at node $i$ in the arborescence.
If node $i$ is not the root of the arborescence, we set the weight of
this arborescence to zero.

\section{Influence of nodes in downstream components}\label{sec:our def}

\subsection{Definition of the extended influence}\label{sub:def of extended influence}

With the definition given by \EQ\eqref{eq:def influence general},
nodes that do not belong to any uppermost SCC have $v_i=0$.  In
practice, however, we often need to assess the relative importance of
different nodes in downstream SCCs and that of nodes in different downstream
SCCs.  There also arise occasions when we want to compare
uninfluential nodes in an uppermost SCC and influential nodes in a
downstream SCC.

An extreme situation in which this is the case is realized by the
network shown in \FIG\ref{fig:toy example N=3}(a). Whenever
$\epsilon>0$ and $\alpha>0$, we obtain $v_1=v_2=0$ and
$v_3=1$. However, when $\epsilon$ and $\alpha$ are small, node 1 may
be regarded to be more central than node 3 because node 1 is much more
central than node 2 and node 3 only weakly influences noncentral node
2.  To cope with such a situation, we extend the influence to
a one-parameter
family of centrality measure by adopting the concept
behind the definition of the
PageRank.

The PageRank of node $i$, denoted by
$R_i$, is defined as the stationary density of the 
discrete-time simple random walk as follows
\cite{Brin98-Langville06book}:
\begin{equation}
R_i=(1-q)\sum_{j=1}^N  \frac{w_{ji}}
{\sum_{\ell=1}^N w_{j\ell}} R_j
+\delta_{\sum_{\ell=1}^N w_{i\ell},0}(1-q)R_i
+ \frac{q}{N}, \quad (1\le i\le N),
\label{eq:page}
\end{equation}
where $\delta_{i,j}=1$ if $i=j$ and $\delta_{i,j}=0$ if
$i\neq j$. 
The so-called teleportation probability $q$
represents the probability that the random walker jumps from any node
to an arbitrary node in one step.
The same concept underlies the definition of the centrality based on the
adjacency matrix \cite{Katz53,Newman10book}.
According to the second term on
the right-hand side in
\EQ\eqref{eq:page}, the random walker stays
at the node without outgoing links with probability $1-q$.
The introduction of $q$ is necessary for treating
networks that are not strongly connected.

The PageRank is originally designed for web graphs. Therefore,
receiving links increases $R_i$, which is opposite to the contention of
the influence. To relate the
PageRank to the influence, we consider the PageRank in the
network generated by reversing all the links of the original network
\cite{mkk09njp}.
We denote this quantity for node $i$ by $R_i^{\rm rev}$, which is
determined by
\begin{equation}
R_i^{\rm rev}=
(1-q)\sum_{j=1}^N  \frac{w_{ij}}
{\sum_{\ell=1}^N w_{\ell j}} R_j^{\rm rev}
+\delta_{\sum_{\ell=1}^N w_{\ell i},0}(1-q)R_i^{\rm rev}
+\frac{q}{N},\quad (1\le i\le N).
\label{eq:page reverse}
\end{equation}

As explained in \SEC\ref{sec:def for general networks},
the influence corresponds to the continuous-time random walk on the
link-reversed network. In a strongly connected network, the influence
of each node is equal to the stationary density of the continuous-time 
random walk on the link-reversed network \cite{MasudaOhtsuki09njp,mkk09njp}.
As in the definition of the PageRank,
let us introduce random global jumps to the continuous-time
random walk on the link-reversed network.
We do so by assuming that
the walker jumps from any node to an arbitrary node with
rate $q$. 
Note that $q$ represents a probability in the PageRank,
whereas it is a rate in the influence. In the following, we allow
$q$ to exceed unity unless otherwise stated.
The destination of the random jump is chosen from all the nodes
with equal probability $1/N$. 
We denote by
$\hat{v}_i$ the stationary density of the modified random walk
at node $i$. The normalization is given by
$\sum_{i=1}^N \hat{v}_i=1$.
The stationary density is obtained from
\begin{equation}
\frac{d \hat{v}_i}{dt}=
\sum_{j=1; j\neq i}^N 
\hat{v}_j w_{ij} - \hat{v}_i \sum_{j=1; j\neq i}^N w_{ji}
+\frac{q}{N} - q  \hat{v}_i = 0.
\label{eq:CTRW with global jumps}
\end{equation}
We define the extended influence by
the solution of \EQ\eqref{eq:CTRW with global jumps}.
We note that the link-reversed version of 
\EQ\eqref{eq:CTRW with global jumps}, with a different 
structure of the global jump, was proposed as an alternative of
the PageRank to be applied to web graphs
\cite{LiuGao08ACM-LiuLiu10InfRetrieval}.

In the vector notation, \EQ\eqref{eq:CTRW with global jumps}
is represented by
\begin{equation}
(\hat{v}_1\; \cdots\; \hat{v}_N) (L+q I)=
\frac{q}{N}(1\; \cdots\; 1),
\label{eq:linear system for mu-influence}
\end{equation}
or equivalently,
\begin{equation}
(\hat{v}_1\; \cdots\; \hat{v}_N) (L+q I-\frac{q}{N}J)=0,
\label{eq:linear system for mu-influence 2}
\end{equation}
where $J$ is the $N$ by $N$ matrix whose all the elements are equal to
unity.
If $q>0$, $L+q I$ is strictly diagonally dominant, and
\EQ\eqref{eq:linear system for mu-influence} can be solved
by the Jacobi or
Gauss-Seidel iteration. A large $q$ guarantees 
exponentially fast convergence of the iteration \cite{Golub96book}.

We note that
\begin{equation}
L+q I = \sum_{m^{\prime}=1}^m q \bm u^{(0)}_{m^{\prime}}
\bm v^{(0)}_{m^{\prime}}
+ (\lambda_2+q) \bm u^{(\lambda_2)}_1 \bm v^{(\lambda_2)}_1 + \ldots,
\end{equation}
which leads to
\begin{equation}
(L+q I)^{-1} = \sum_{m^{\prime}=1}^m \frac{1}{q} \bm u^{(0)}_{m^{\prime}}\bm v^{(0)}_{m^{\prime}}
+ \frac{1}{\lambda_2+q} \bm u^{(\lambda_2)}_1 \bm v^{(\lambda_2)}_1 + \ldots.
\label{eq:(L+mu I)^{-1}}
\end{equation}
In the limit $q\to 0$, \EQ\eqref{eq:(L+mu I)^{-1}} implies that
$\hat{v}_i\approx v_i$, where $v_i$ is defined by
\EQ\eqref{eq:def of extended v_i}.
For the first term on the right-hand side of
\EQ\eqref{eq:def of extended v_i} to be comparable with the remaining terms,
$q$ must be at least
approximately ${\rm Re}\lambda_2$. If this is the case, $\hat{v}_i$ can
quantitatively differentiate various nodes including those
in downstream components.
 
In the limit $q\to\infty$,
\EQ\eqref{eq:linear system for mu-influence} gives
$\hat{v}_i=1/N$ ($1\le i\le N$).  When $q$
is a large finite value, \EQ\eqref{eq:linear system for mu-influence}
is expanded as
\begin{align}
(\hat{v}_1\; \cdots\; \hat{v}_N) = & \frac{1}{N}(1\; \cdots\; 1)
\sum_{\ell=0}^{\infty}(-1)^{\ell}\left(\frac{L}{q}\right)^\ell\notag\\
=& \frac{1}{N}(1\; \cdots\; 1)+\frac{1}{Nq}
(k_1^{\rm out}-k_1^{\rm in}\; \cdots\; k_N^{\rm out}-k_N^{\rm in})
+O\left(\frac{1}{q^2}\right),
\label{eq:Taylor}
\end{align}
where $k_i^{\rm out}$ and $k_i^{\rm in}$ are the outdegree and the
indegree of node $i$, respectively.  The Taylor expansion is justified
when $q>{\rm Re}\lambda_N$, where $\lambda_N$ is the eigenvalue of $L$
with the largest modulus.  If $q$ is large relative to ${\rm
  Re}\lambda_N$, the influence is determined by the outdegree and the
indegree and is independent of the global structure of networks. Therefore,
in practice, $q$ should not be too large as compared to ${\rm
  Re}\lambda_N$.  This is surprising because a large $q$ implies a
strong global connectivity.  As a rule of thumb, we recommend setting
${\rm Re}\lambda_2 < q < {\rm Re}\lambda_N$.  A suitable range of the
teleportation probability $q$ for the PageRank can be also obtained by
applying the criterion ${\rm Re}\lambda_2 < q < {\rm Re}\lambda_N$ to
the PageRank matrix implied in \EQ\eqref{eq:page}.

\subsection{Interpreting the extended influence without regard to global jumps}\label{sub:relation to PageRank}

We have extended the influence by introducing global jumps to the
continuous-time random walk on the link-reversed network. However, the
meaning of the teleportation term in terms of the dynamical and structural
properties of the nodes in the network and its rationale are
somewhat vague. We show that the extended influence defined by
\EQ\eqref{eq:linear system for mu-influence} allows another
interpretation: absorption probability of the random walk on the
link-reversed network with a sink attached to each node but without
global jumps. 
A similar interpretation was made for the PageRank in
Ref.~\cite{Avrachenkov2007siam}.

\begin{figure}
\begin{center}
\includegraphics[width=8cm]{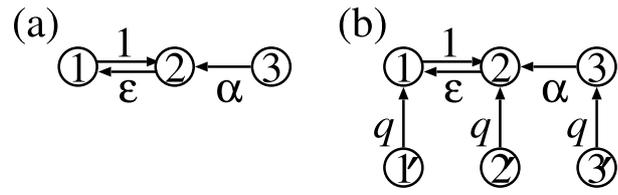}
\caption{(a) Example of directed network with $N=3$. (b)
Network with $2N=6$ nodes obtained by adding source nodes to the
network shown in (a).}
\label{fig:toy example N=3}
\end{center}
\end{figure}

We assume $N$ additional source nodes indexed by $1^{\prime}$, $\ldots$,
$N^{\prime}$ and directed links with weight $q>0$ from
node $i^{\prime}$ to node $i$ ($1\le i\le N$) in the original network. 
The extension of the network shown in
\FIG\ref{fig:toy example N=3}(a) is depicted in 
\FIG\ref{fig:toy example N=3}(b).
The extended network has $2N$ nodes.
Nodes  $1^{\prime}$, $\ldots$, $N^{\prime}$ are the unique root nodes
of the extended network. Node $i^{\prime}$ forms the $i$th uppermost
SCC in the extended network. The multiplicity of the zero Laplacian
eigenvalue of the extended network is equal to $m=N$.

We then reverse all the links and consider
the probability that the random walker starting from an arbitrary node
with equal probability $1/2N$
is absorbed at node $i^{\prime}$. This probability is given by
$v_{i^{\prime}}$. Because it is obvious and uninformative that
the random walker starting from the auxiliary node $i^{\prime}$ is
necessarily absorbed to node $i^{\prime}$, we would like to exclude
this factor. Therefore, we examine the quantity given by
\begin{equation}
2\left(v_{i^{\prime}}-\frac{1}{2N}\right)=2
v_{i^{\prime}}-\frac{1}{N}.
\label{eq:mu-influence def}
\end{equation}
The subtraction of $1/(2N)$ in \EQ\eqref{eq:mu-influence def}
accounts for the exclusion of the random
walker starting from and absorbed to node $i^{\prime}$.
The multiplicative factor 2 accounts for the fact that we effectively
start the random walk from nodes 1, $\ldots$, $N$ with equal probability
$1/N$.

Equation~\eqref{eq:def influence general} implies that
the calculation of $v_{i^{\prime}}$ involves
$v_{i,1}$, that is, the first element of the left zero eigenvector of
$L$ corresponding to the $i$th uppermost SCC. Because the uppermost SCC
consists of single node $i^{\prime}$, $v_{i,1}$ is
equal to unity.
The calculation of $v_{i^{\prime}}$ also involves 
$\bm u^{(0)}_i\equiv (\bm e_i\; \overline{\bm u}_i)^{\top}$, that is,
the zero eigenvector of
the $2N$-dimensional Laplacian.
$\overline{\bm u}_i^{\top}$ is an $N$-dimensional column vector.
By substituting these expressions and
\EQ\eqref{eq:def influence general} in \EQ\eqref{eq:mu-influence def},
the quantity given by \EQ\eqref{eq:mu-influence def} is equal to
\begin{equation}
2\left(v_{i,1}\frac{\sum_{j=1}^{2N} u^{(0)}_{i,j}}{2N}
-\frac{1}{2N}\right) = \frac{\sum_{j=1}^N \overline{u}_{i,j}}{N}.
\label{eq:mu-influence concise expression}
\end{equation}
We calculate $\overline{\bm u}_i^{\top}$ from
\begin{equation}
\left(\begin{array}{cc}
O&O\\
-q I& L+q I
\end{array}\right)
\left(\begin{array}{c}
\bm e_i \\ \overline{\bm u}_i
\end{array}\right)=0,
\label{eq:2N eigen eqn}
\end{equation}
where $O$ is the $N \times N$ zero matrix and
$L$ is the Laplacian of the original network.
Equation~\eqref{eq:2N eigen eqn} is equivalent to
\begin{equation}
(L+q I)\overline{\bm u}_i=q \bm e_i.
\label{eq:2N->N eigen eqn}
\end{equation}
By combining \EQS\eqref{eq:mu-influence concise expression} and \eqref{eq:2N->N eigen eqn},
we obtain
\EQ\eqref{eq:linear system for mu-influence}.

With this interpretation, we gain an intuitive understanding 
of the fact that the extended influence $\hat{v}_i$
is a
local quantity when $q$ is large.  In this
situation, the tendency that a random walk exits from each node is
strong, and a random walk would not travel a long distance before
being absorbed.  Therefore, it is natural that $\hat{v}_i$ at large
$q$ is efficiently approximated by local quantities of
nodes such as the outdegree and the indegree, 
as discussed using \EQ\eqref{eq:Taylor}.

\section{Toy examples}\label{sec:toy examples}

In this and the next section, 
we apply the extended influence to various networks.

\subsection{Network with $N=3$}

Consider the network shown in
\FIG\ref{fig:toy example N=3}(a).  We are concerned with the situation
in which $0< \epsilon\ll 1$ such that node 1 is apparently
much more central than node 2. 
If node 3 is absent, $v_1/v_2=1/\epsilon$;
node 1 is actually much more influential than node 2
\cite{kori09,mkk09pre}. However, regardless of the value of
$\alpha>0$, node 3 takes all the share of the influence if we use
$v_i$.

The extended influence $\hat{v}_i$ ($1\le i\le 3$) is equal to
$v_i$ for the network shown in \FIG\ref{fig:toy example N=3}(b).
We obtain
\begin{align}
\hat{v}_1 =& \frac{1}{3\Delta}\left[q^2+\left( 2+\alpha\right)q\right],\\
\hat{v}_2 =& \frac{1}{3\Delta}\left(q^2 + 2\epsilon q\right),\\
\hat{v}_3 =& \frac{1}{3\Delta}\left[
q^2 + \left(1+\epsilon+2\alpha\right)q + 3\epsilon\alpha\right],
\end{align}
where
\begin{equation}
\Delta = q^2 + (1+\epsilon+\alpha)q + \epsilon\alpha.
\end{equation}
Therefore, $\hat{v}_1>\hat{v}_3$ when $(1-\epsilon-\alpha)q>3\epsilon\alpha$.
When $\epsilon$ or $\alpha$ is small and
$\epsilon+\alpha<1$, we have an intuitive result that node 1 
is more influential than node 3.

\subsection{Directed chain}\label{sub:directed chain}

Consider a directed chain having $N$ nodes
defined by $w_{i,i+1}=1$ ($1\le i\le N-1$) and $w_{i,j}=0$ ($j\neq i+1$).
The network is schematically shown
in \FIG\ref{fig:directed chain}.
We obtain $v_1=1$ and $v_i=0$ ($2\le i\le N$)
\cite{mkk09pre}. However, nodes with small $i$
are located relatively upstream in the chain and
intuitively appear influential as compared to nodes
with large $i$. We can calculate
the influence either by
solving \EQ\eqref{eq:2N->N eigen eqn} or by analyzing random walks
with $N$ traps on the network obtained by reversing all the links
shown in \FIG\ref{fig:directed chain}. When the random walker
on the link-reversed network starts from node $j$ ($2\le j\le N$),
the probability that the walker exits from node $i$
to the absorbing node $i^{\prime}$ is equal to
\begin{equation}
u^{(0)}_{ij}=\begin{cases}
\frac{1}{(1+q)^{j-1}}, & (i=1),\\
\frac{q}{(1+q)^{j-i+1}}, & (2\le i\le j),\\
0, & (j+1\le i\le N).
\end{cases}
\label{eq:p_ij for chain}
\end{equation}
Therefore, we obtain
\begin{equation}
\hat{v}_i = \frac{1}{N}\sum_{j=1}^N u^{(0)}_{ij}=
\begin{cases}
\frac{1}{N}\left(\frac{1+q}{q}-\frac{1}{q(1+q)^{N-1}}\right), & (i=1),\\
\frac{1}{N}\left(1-\frac{1}{(1+q)^{N-i+1}}\right), & (2\le i\le N).
\end{cases}
\label{eq:hat-vi chain}
\end{equation}
We note that $\lim_{q\to 0} \hat{v}_1=1$,
$\lim_{q\to 0}\hat{v}_i=0$ 
($2\le i\le N$), and $\hat{v}_i$ monotonically decreases with $i$
for any $q>0$.

\section{Numerical results}\label{sec:large-N examples}

In this section, we examine the influence in three directed networks:
a random graph,
a neural network, and an online social network.

\subsection{Descriptions of networks}

We generate a directed random network with $N=100$ and expected degree
$\left<k\right>=3.5$ by
connecting each ordered pair of nodes independently with
probability $\left<k\right>/(N-1)$. Because $\left<k\right>$
is relatively small, the generated network is
not strongly connected, whereas 
it is weakly connected, i.e., not divided into disconnected components.
The generated network has three root nodes, each of which 
forms an SCC. The largest SCC contains 94 nodes and
is downstream to the three root nodes. The extremal Laplacian eigenvalues
are $\lambda_2=0.046$ and $\lambda_N=8.255$.

We generate a \textit{C.~elegans} neural network with $N=279$ on the
basis of published data \cite{Chen06pnas-wormatlas}.  In this
network, there exist two types of links: undirected gap junctions and
directed chemical synapses.  A pair of neurons can be connected by
multiple synapses.  We regard this network as a weighted directed
network, where the weight of the link from neuron $i$ to neuron $j$ is
defined as the summation of the number of gap junctions between
$i$ and $j$ and the number of chemical synapses from $i$ to $j$.
The network has 2993 links.  The largest SCC has 274 nodes
\cite{mkk09njp}. Four of the five remaining nodes are located upstream
to the largest SCC and form individual SCCs.
The other node is located downstream to
the largest SCC. The extremal Laplacian eigenvalues are
$\lambda_2=0.050$ and $\lambda_N=354.105$.

The third network that we use is an online social network  
among students at University of California,
Irvine \cite{Panzarasa09JASIST}.
This network has $N=1899$ nodes and 20296 directed and weighted links.
We focus on the
largest weakly connected component of this network that contains
1893 nodes and 13835 links.
There exist 103 root nodes, each of which forms an SCC.
The largest SCC has 1023 nodes and is downstream to these root nodes.
The extremal Laplacian eigenvalues are
$\lambda_2=0.146$ and $\lambda_N=92.996$.

\subsection{Analysis of influence in the three networks}

\begin{figure}
\begin{center}
\includegraphics[width=8cm]{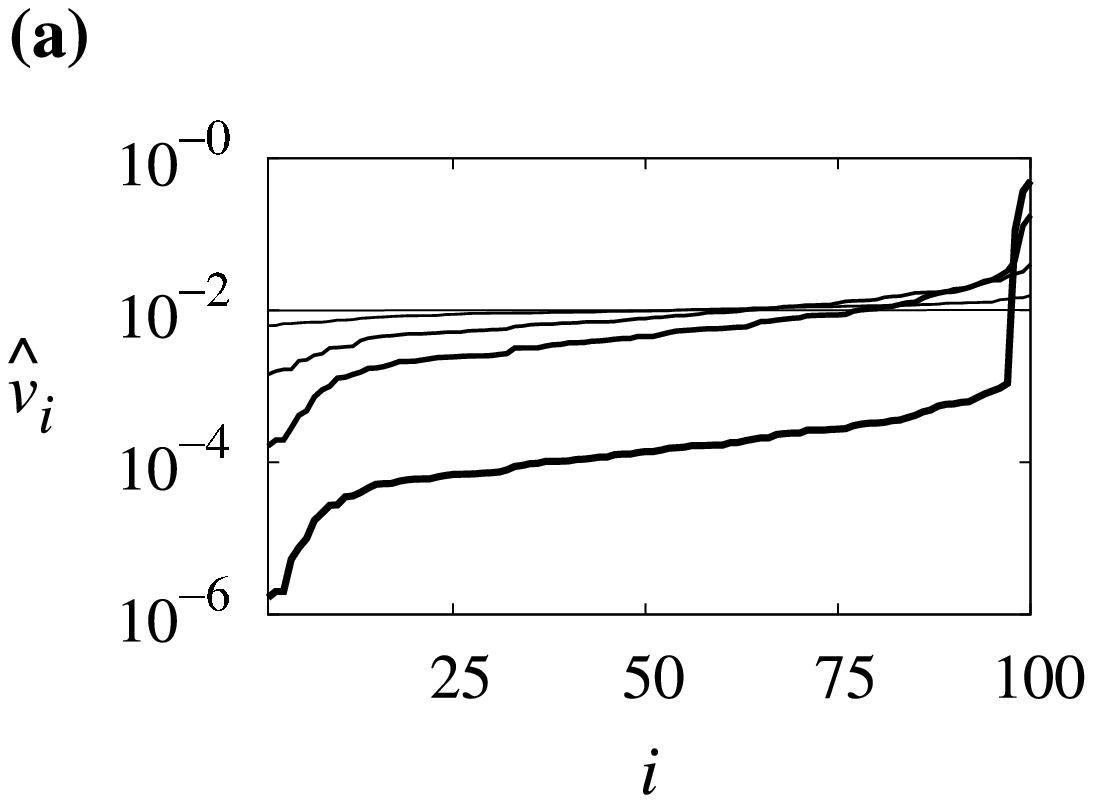}
\includegraphics[width=8cm]{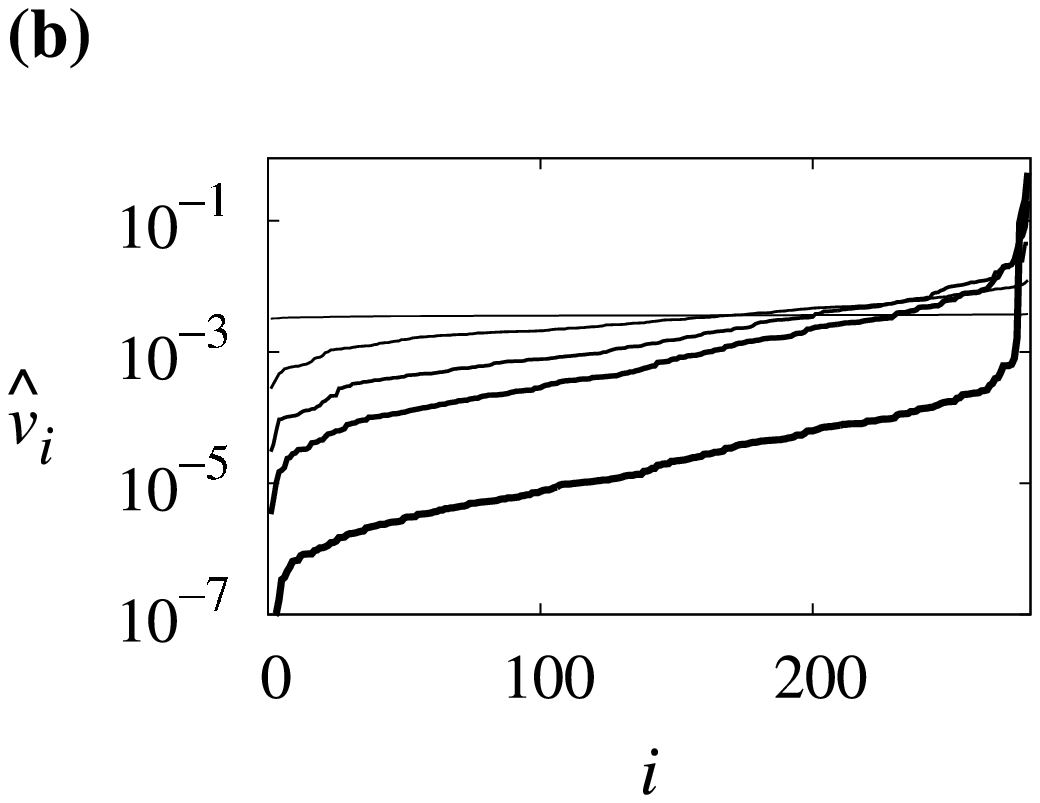}
\includegraphics[width=8cm]{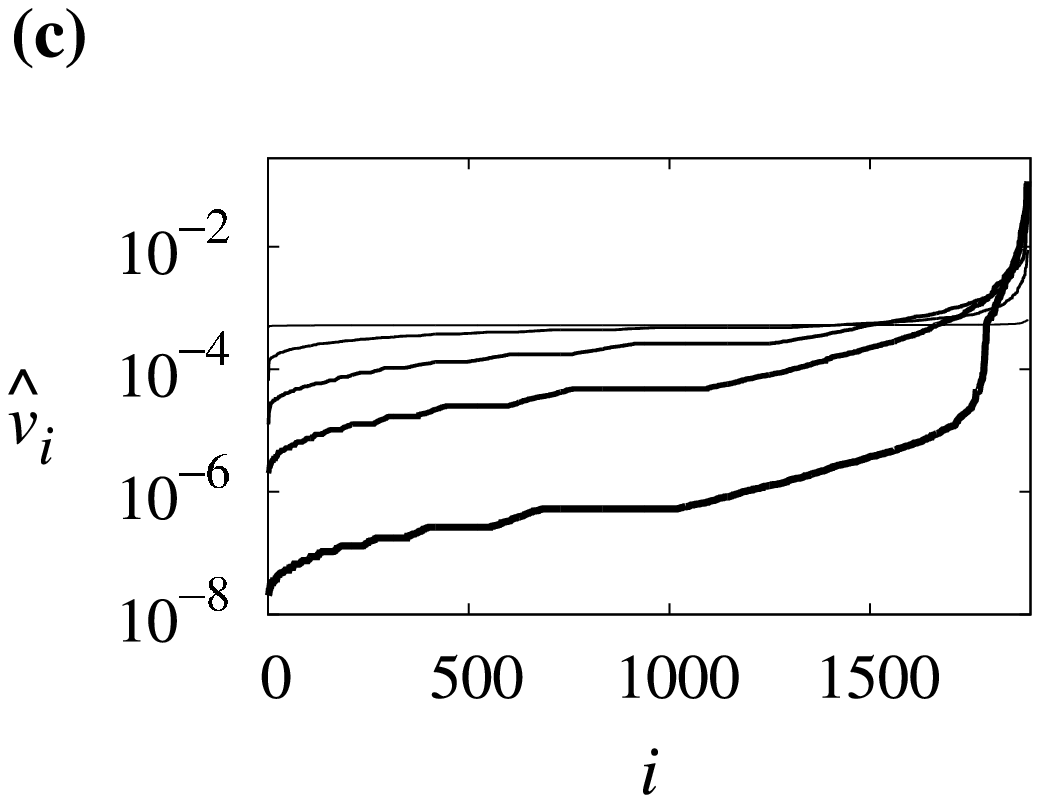}
\caption{Values of $\hat{v}_i$ for (a) random graph, (b)
neural network, and (c) online social network.
We set $q=0.001$, 0.1, 1, 10, 1000 (from steep thick
lines to flat thin lines).
For each $q$, we have sorted 
$\hat{v}_i$ in the ascending order for demonstration.}
\label{fig:ascend}
\end{center}
\end{figure}

The rank plots of the influence
for various values of $q$ 
for the random graph, neural network, and online social
network are shown in 
\FIGS\ref{fig:ascend}(a)--(c), respectively.
In the figure, the values of $\hat{v}_i$ 
are shown in the ascending order for each $q$
for clarity. 

When $q=0.001$ (thickest lines),
 $\hat{v}_i$ is similar to
$v_i$ for the three networks. Therefore,
the root nodes have exclusively large $\hat{v}_i$, whereas the other nodes have
$\hat{v}_i\approx 0$. Accordingly, we find
a sudden jump in the rank plot for each network.
Such a small value of $q$ does not allow us to
quantitatively compare the centrality of nodes in downstream components.
This is also anticipated from the fact that the three networks yield
$q =0.001 < {\rm Re}\lambda_2$.
In the other extreme, $\hat{v}_i\approx 1/N$ is roughly satisfied
when $q=1000$ (thinnest lines).
This is consistent with the fact that the three networks yield
$q=1000> {\rm Re}\lambda_N$.
In this range of $q$,
the influence is not an adequate centrality measure.
For intermediate values of $q$,
$\hat{v}_i$ is reasonably dispersed, and nodes that are not the
roots are also endowed with positive $\hat{v}_i$. We consider that
the influence with intermediate values of $q$ enables us to
compare the importance of nodes that are in downstream SCCs and
quantify the relative importance of nodes in uppermost SCCs
and nodes in downstream SCCs.

The influence with intermediate values of $q$ is 
distinct from the interpolation of the influence when $q\to 0$
(i.e., $\hat{v}_i\approx v_i$) and that when $q=\infty$ (i.e., $\hat{v}_i=1/N$).
The order of the nodes in terms of the value of $\hat{v}_i$
drastically changes as $q$ varies.
To demonstrate this, we examine the dependence of $\hat{v}_i$
on $q$ for some selected nodes.

\begin{figure}
\begin{center}
\includegraphics[width=8cm]{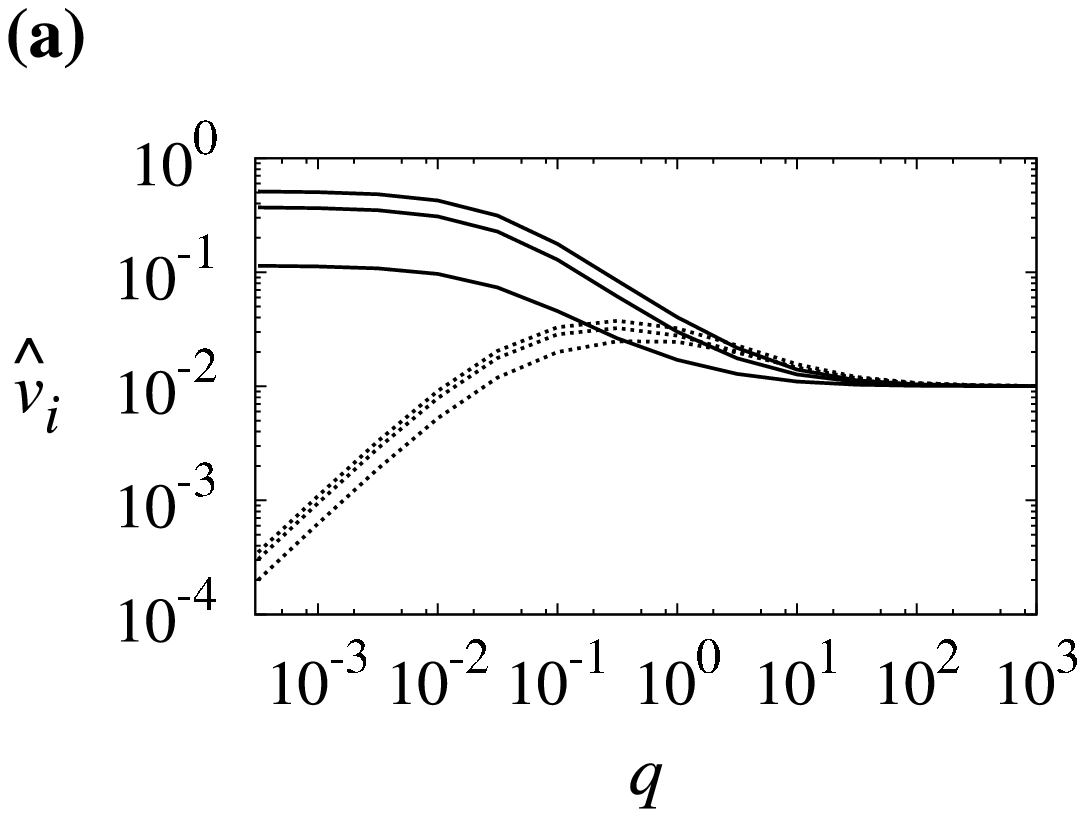}
\includegraphics[width=8cm]{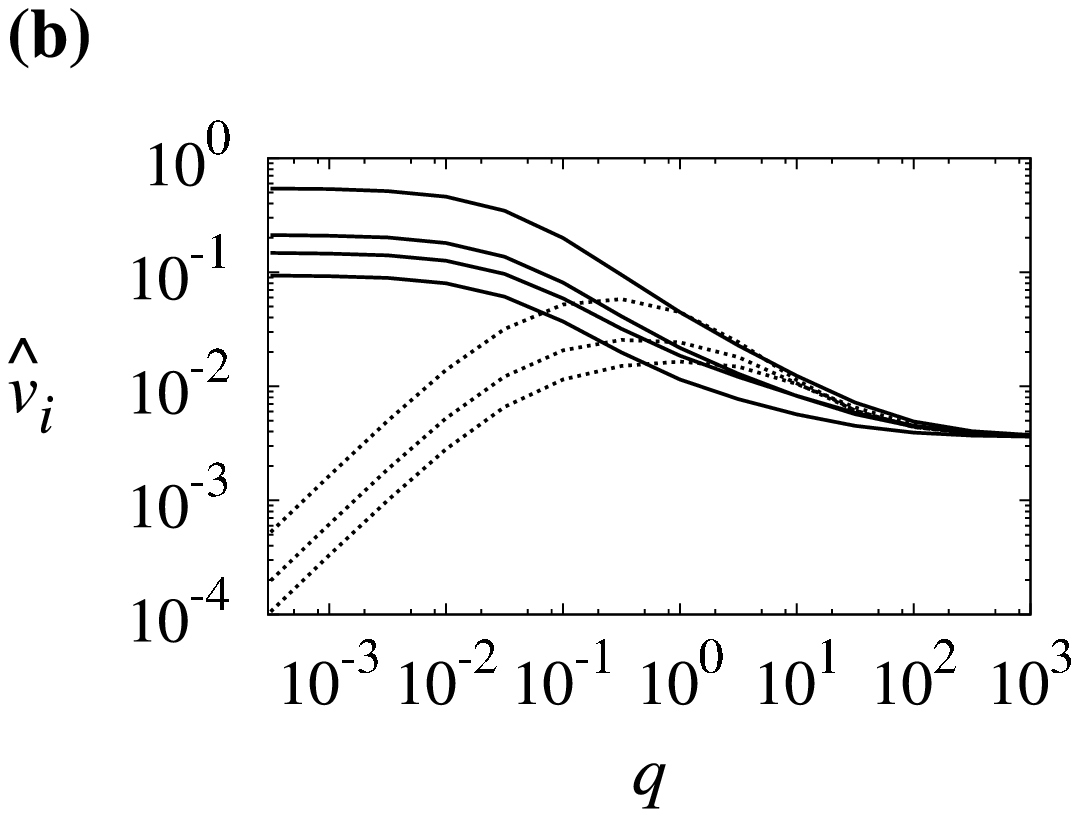}
\includegraphics[width=8cm]{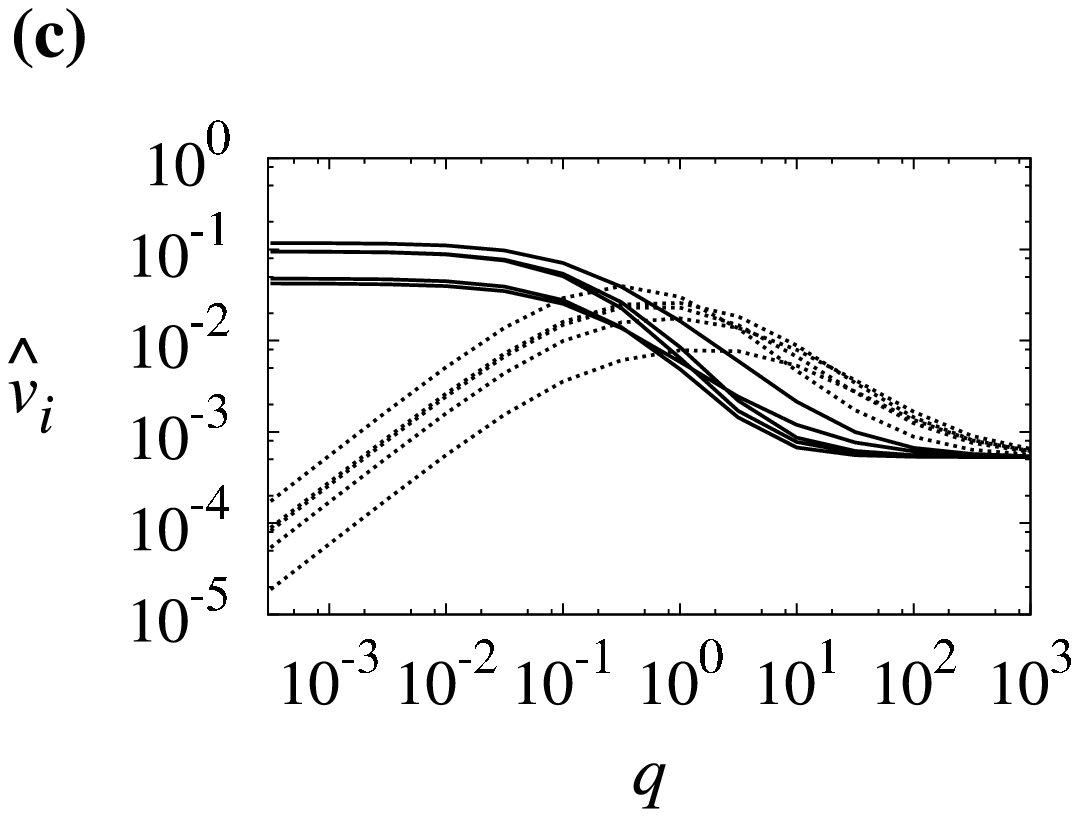}
\caption{Dependence of $\hat{v}_i$ of some
nodes on $q$ for (a) random graph, (b)
neural network, and (c) online social network. The nodes with the
largest influence values for $q=0.001$ and for $q=10$ correspond
to the solid and dashed lines, respectively.}
\label{fig:vary q}
\end{center}
\end{figure}

For the random graph, we select the three root nodes, for which
$\hat{v}_i$ is the largest at $q=0.001$ and the three nodes whose
$\hat{v}_i$ is the largest at $q=10$. The dependence of $\hat{v}_i$
on $q$ for the six nodes is shown in \FIG\ref{fig:vary q}(a). The
three root nodes [solid lines in \FIG\ref{fig:vary q}(a)]
and the three nodes with the largest $\hat{v}_i$ at
$q=10$ (dashed lines)
do not overlap each other.  In particular, the root node with
the third largest $\hat{v}_i$ for $q=0.001$ does not have large
$\hat{v}_i$ when $q$ is approximately larger than 1. Although the
indegree of this root node is equal to zero, the destinations of the
links from this root node are presumably nodes with small influence
values in the largest SCC. This phenomenon is essentially the same as
that shown in \FIG\ref{fig:toy example N=3}.

The neural network has four root nodes. The dependence of
$\hat{v}_i$ on $q$ for the root nodes and the three nodes whose 
$\hat{v}_i$ is among the
four largest values at $q=10$ are shown in
\FIG\ref{fig:vary q}(b). In the neural network,
one of the four roots is among
the nodes with the four largest values of $\hat{v}_i$ at $q=10$. 
For the online social network,
the relationships between $\hat{v}_i$ and $q$
for the five root nodes with the largest $\hat{v}_i$ at $q=0.001$
and the five nodes with the largest $\hat{v}_i$ at $q=1000$
are shown in \FIG\ref{fig:vary q}(c).
The results for the neural network and the online social network
are qualitatively the same as those for the random graph.
In particular, some root nodes (solid lines)
do not have particularly large $\hat{v}_i$ when
$q$ is approximately larger than unity.

\begin{figure}
\begin{center}
\includegraphics[width=5cm]{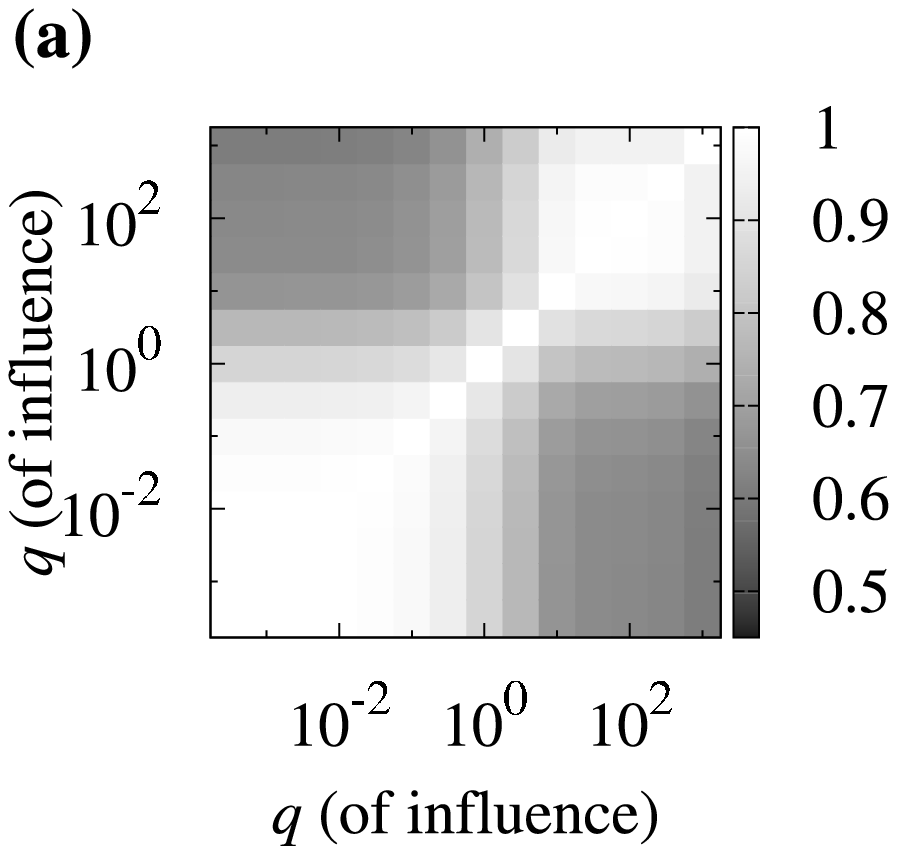}
\includegraphics[width=5cm]{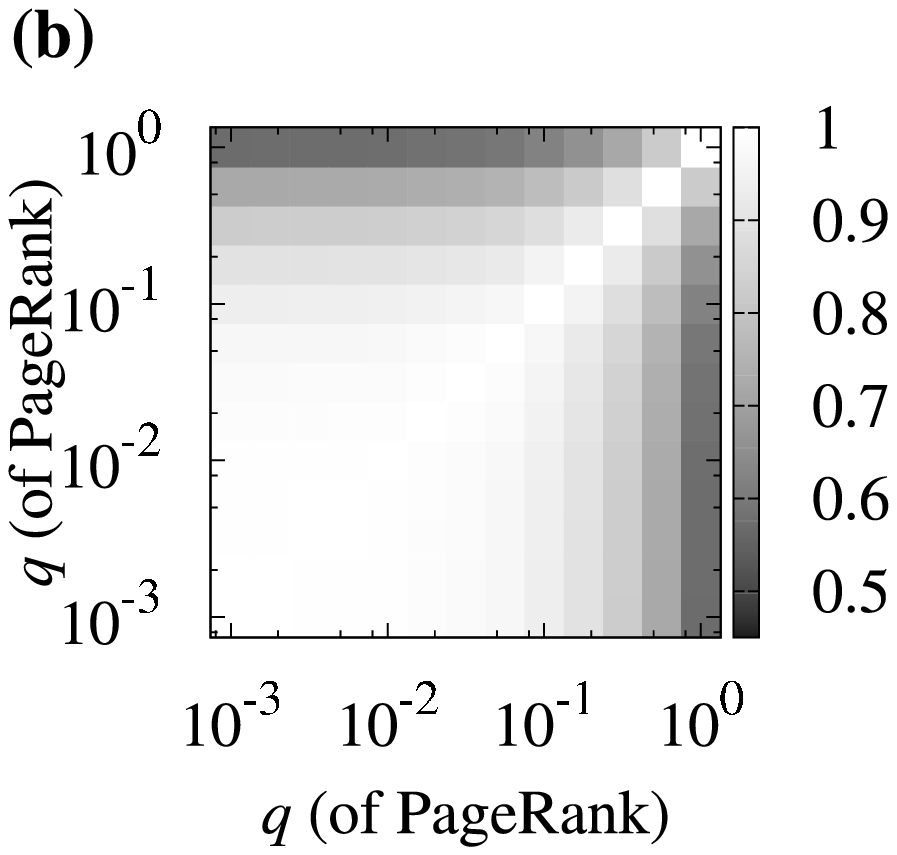}
\includegraphics[width=5cm]{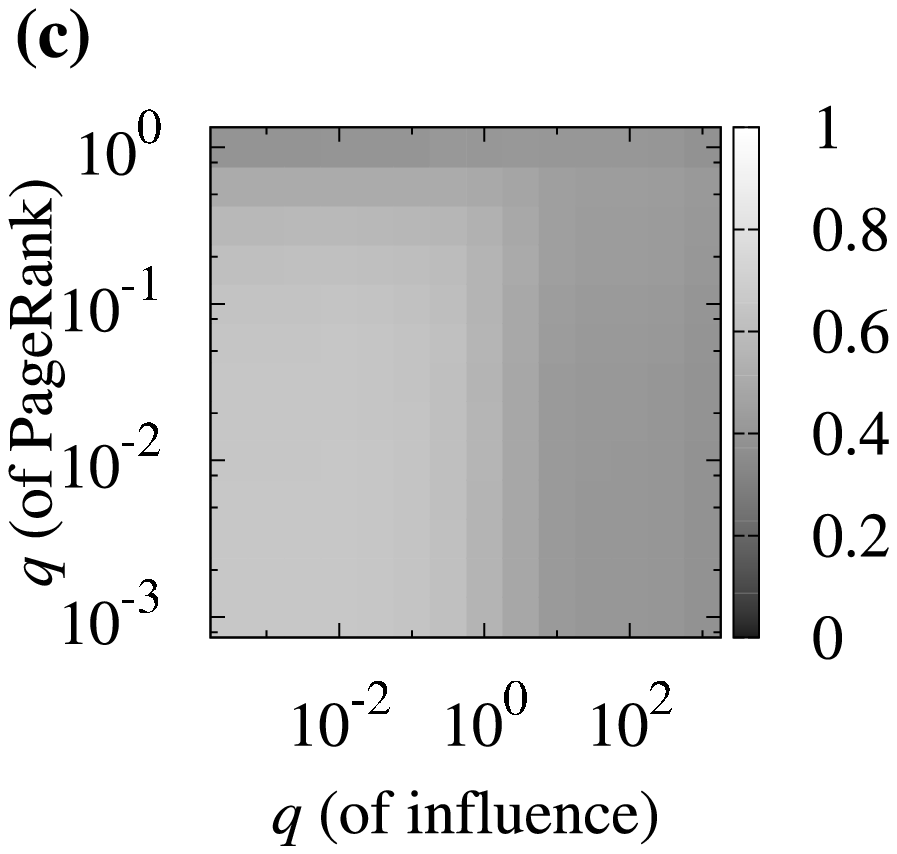}
\includegraphics[width=5cm]{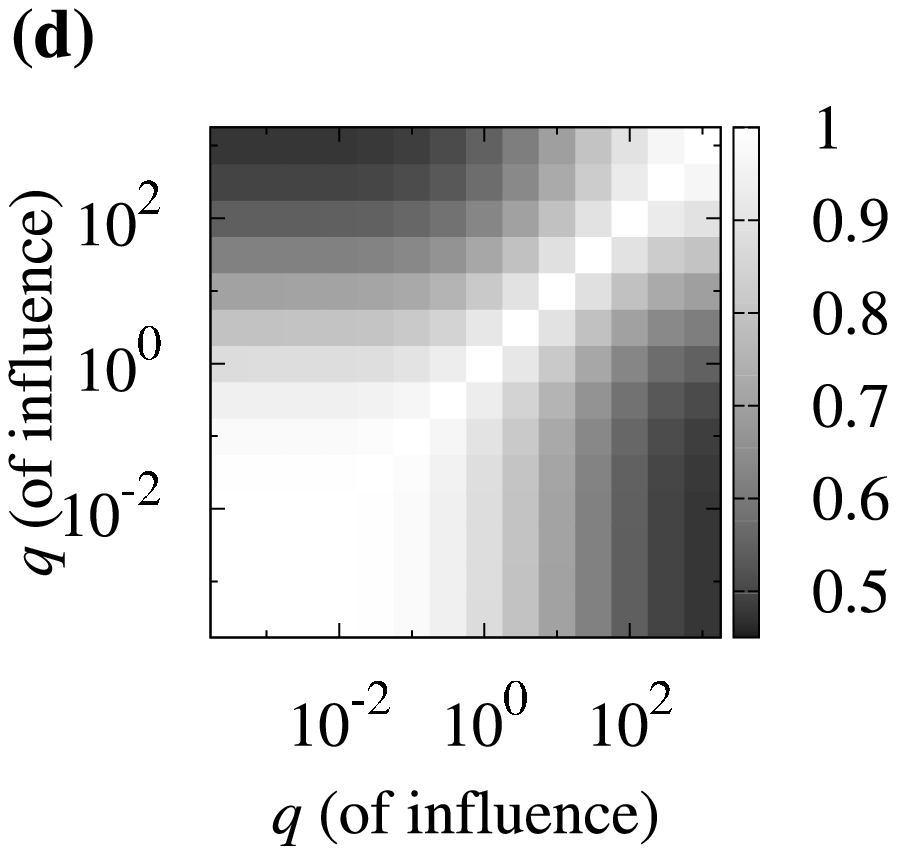}
\includegraphics[width=5cm]{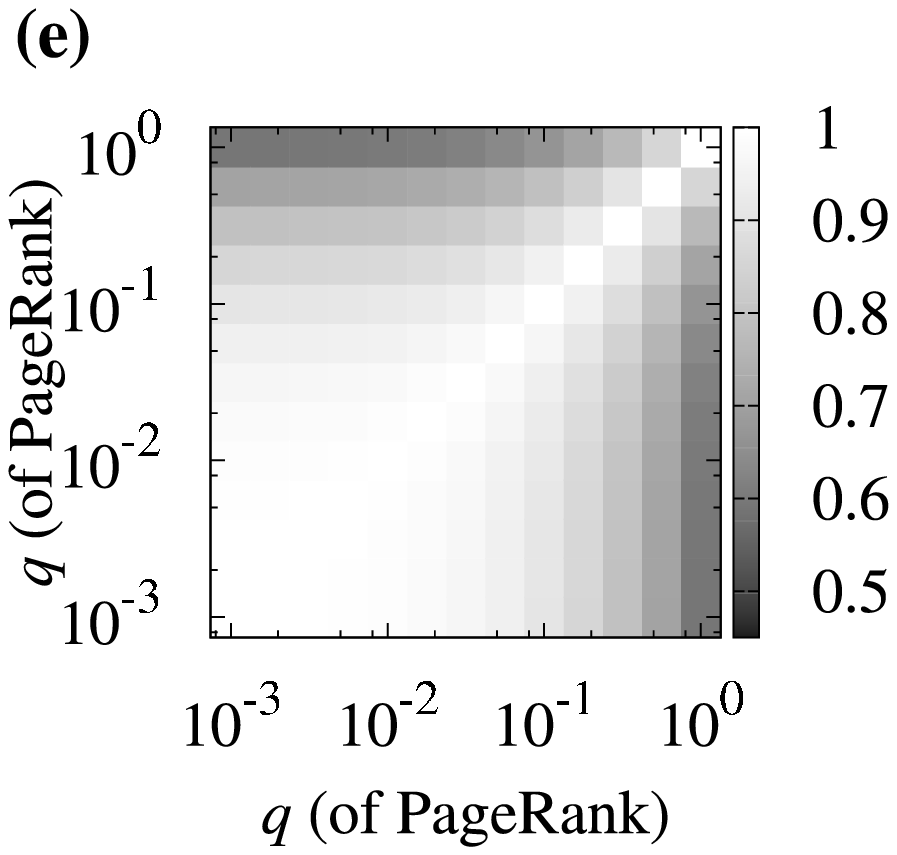}
\includegraphics[width=5cm]{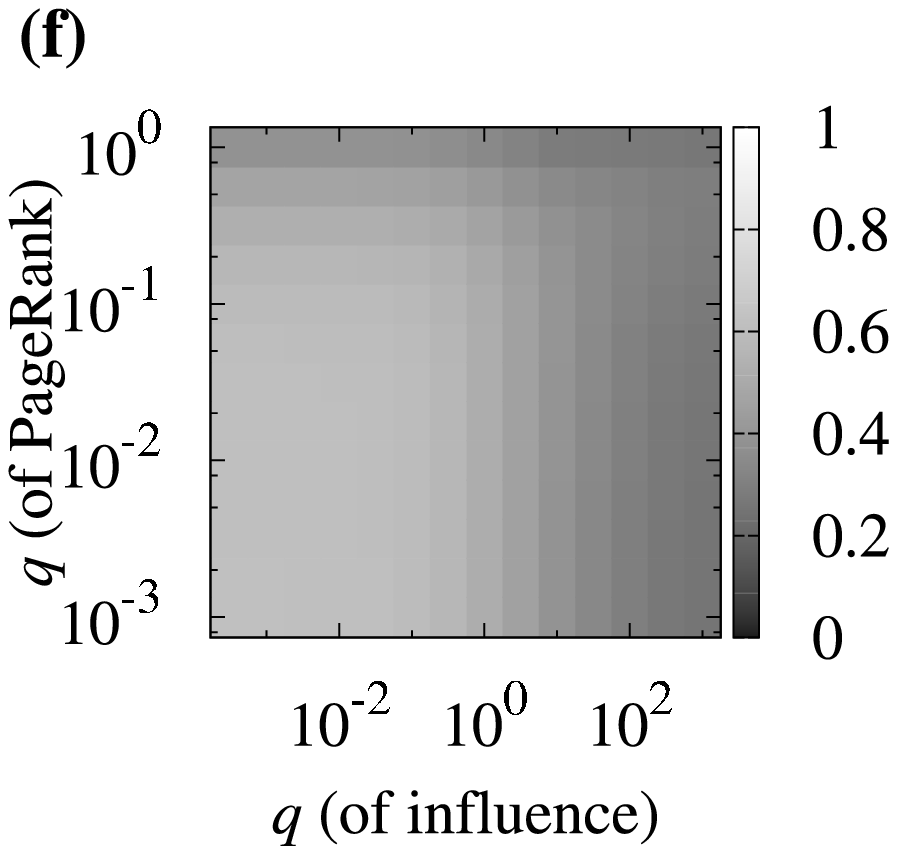}
\includegraphics[width=5cm]{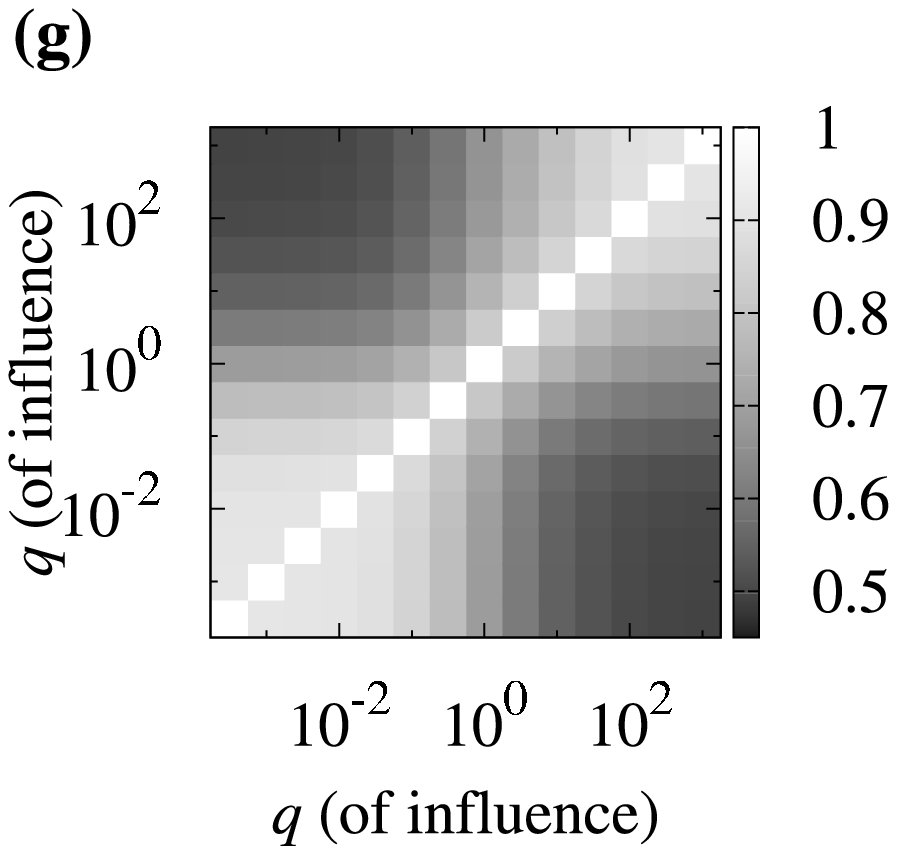}
\includegraphics[width=5cm]{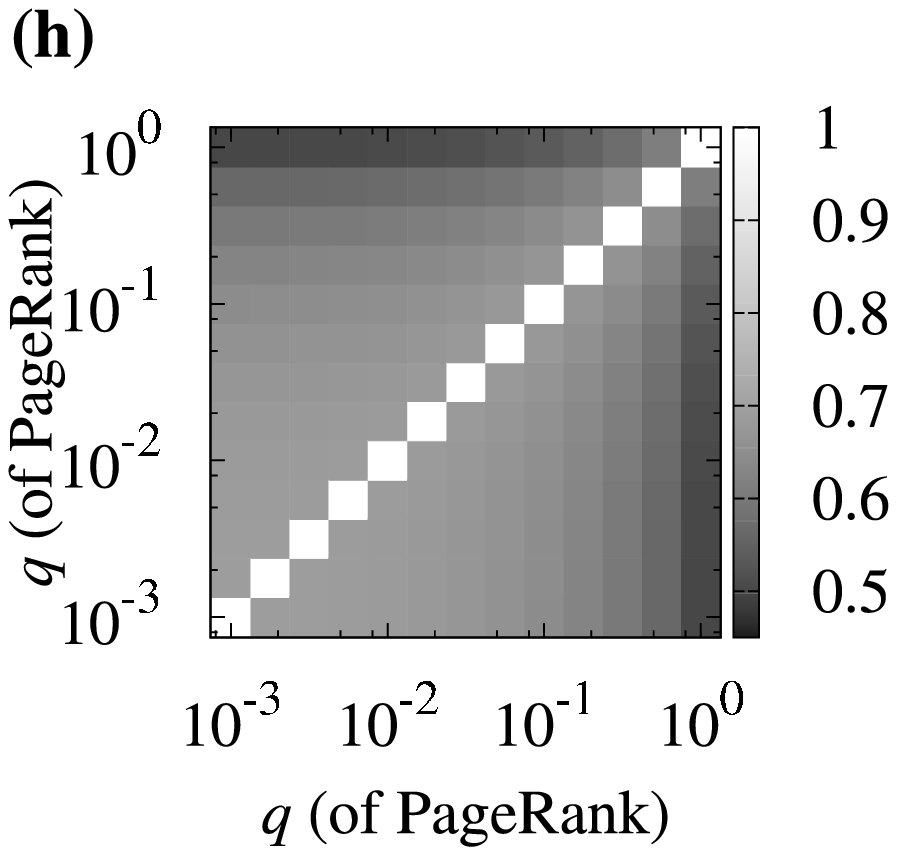}
\includegraphics[width=5cm]{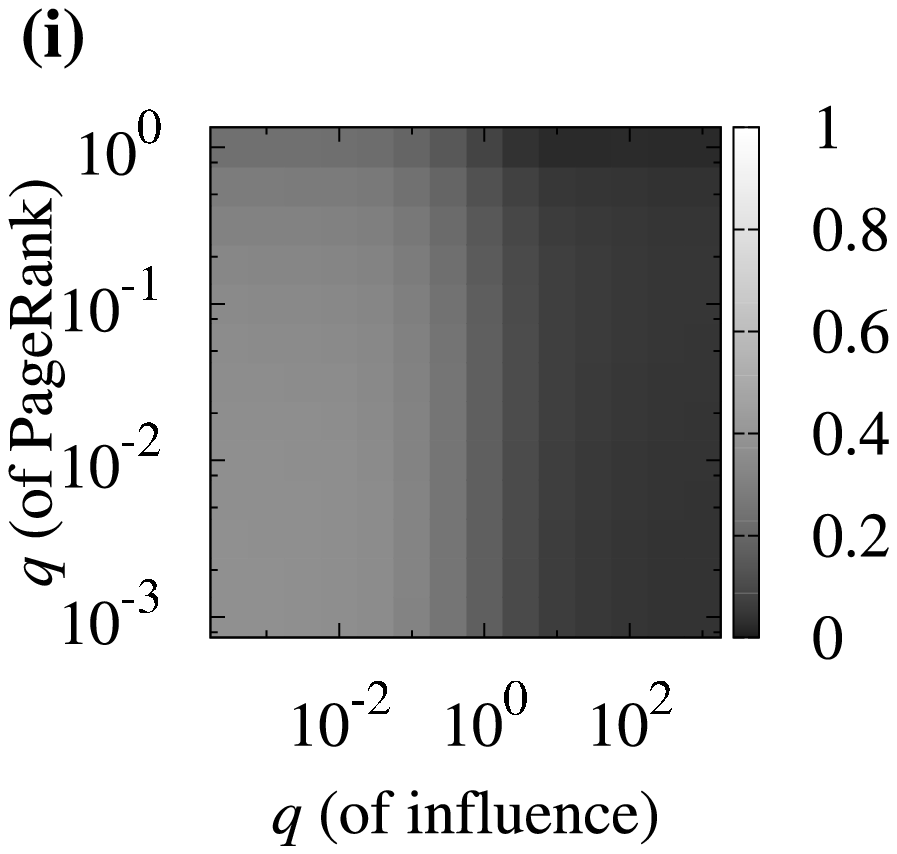}
\caption{Kendall rank correlation coefficient for the influence and
the PageRank.
(a)--(c) Random graph, (d)--(f) neural network, and (g)--(i) online social network.
(a, d, g) Correlation between the influence for different values of $q$.
(b, e, h) Correlation between the PageRank for different values of $q$.
(c, f, i) Correlation between the influence and the PageRank for various values of $q$.}
\label{fig:influence vs PageRank}
\end{center}
\end{figure}

Finally, we quantify the dependence of the influence on 
$q$ by calculating the Kendall rank correlation coefficient.
It is defined as $2\{P/[N(N-1)/2]-1\}$, where $P$ is the number of pairs $i$, $j$
($1\le i < j \le N$) such that the sign of $\hat{v}_i-\hat{v}_j$ for
$q=q_1$ is the same as that for $q=q_2$. The correlation coefficient falls between $-1$ and $1$.
The correlation coefficient for the random graph for various values of
$q$ is shown in 
\FIG\ref{fig:influence vs PageRank}(a). As anticipated,
the correlation decreases with $|q_1-q_2|$. Figure~\ref{fig:influence vs PageRank}(a)
also indicates that the
ranking on the basis of the influence is fairly insensitive to $q$ in two ranges of
$q$, i.e., for $q$ smaller than
$\approx 1$ and for $q$ larger than $\approx 10$. The ranking
is sensitive to $q$ between these two ranges of $q$.
For comparison, the correlation coefficient for the PageRank for various
values of $q$ is shown in \FIG\ref{fig:influence vs PageRank}(b).
Similar to the case of the influence, the correlation decreases with $|q_1-q_2|$.
The correlation between the influence and the PageRank  
[\FIG\ref{fig:influence vs PageRank}(c)] is generally small
regardless of the two values of $q$. On this basis, we claim that the influence
and the PageRank are distinct centrality measures. This result 
generalizes
that when directed networks are strongly connected and $q=0$ \cite{mkk09njp}.

The rank correlation coefficient for the neural network and the
social network calculated in the same manner is shown in 
\FIGS\ref{fig:influence vs PageRank}(d)--(f) and \FIGS\ref{fig:influence
  vs PageRank}(g)--(i),
respectively. The results are qualitatively the same as those for the
random graph (\FIGS\ref{fig:influence vs PageRank}(a)--(c)).

\section{Conclusions}\label{sec:conclusions}

We have proposed a centrality measure (influence) for general directed
networks.  It is a generalization of a Laplacian-based centrality
measure that is often used for strongly connected networks
\cite{Daniels69Biom-Berman80SIAM,Moon70SiamRev,Biggs97BullLondMathSoc,mkk09njp,mkk09pre,Klemm10arxiv}.
It also generalizes the formulation of the same centrality measure
developed for networks that are not necessarily strongly connected
\cite{Agaev00,Borm02,Chebotarev05arxiv}. Unlike the previous
measure \cite{Agaev00,Borm02,Chebotarev05arxiv},
the proposed measure is suitable for comparing the importance
of nodes that are in downstream SCCs and comparing nodes in different
SCCs. It has a free parameter $q$. For networks
that are not strongly connected, we
suggest using ${\rm Re}\lambda_2< q< {\rm Re}\lambda_N$
(\SEC\ref{sub:def of extended influence}).
A small value of $q$ implies that the centrality values concentrate
on nodes in uppermost components. A large value of $q$ makes the influence
close to a degree centrality, i.e., outdegree minus indegree. The choice of
$q$ is up to users' preferences.
We acknowledge that various
mathematical properties of the matrix associated with the
influence (i.e., $L+q I$) have been analyzed in
\cite{Agaev00,Chebotarev05arxiv}. In \cite{Chebotarev05arxiv}, the use of this
matrix for the centrality measure is briefly mentioned.

Arguably, the most frequently used centrality measure for directed
networks appears to be the PageRank
\cite{Brin98-Langville06book}.  Beyond the World Wide
Web, for which the PageRank was originally designed, the PageRank has been
applied to rank, for example, academic papers and journals (e.g.,
\cite{Palacios04-Fersht09}).  The PageRank is interpreted as the
stationary density of the discrete-time simple random walk with global
jumps on the network. 

We have defined the influence as
a continuous-time counterpart of the PageRank.
Furthermore, we have provided the interpretation of the influence 
as the absorption probability of the continuous-time random walk
to the sink attached to each node but not with global random jumps.
As a corollary, the PageRank
can be interpreted as the absorption probability of the random walk
without teleportation to a sink. In addition,
a suitable range of the teleportation probability in the PageRank
can be estimated by adapting
the criterion
${\rm Re}\lambda_2< q< {\rm Re}\lambda_N$
to the discrete-time random walk.

For the case of strongly connected networks, 
we refer to our previous work \cite{MasudaOhtsuki09njp,mkk09njp}
for a discussion of continuous-time versus discrete-time random walk.
We have shown that $q$ controls the
relative importance
of nodes in upstream SCCs and nodes in downstream SCCs. 
The same role is shared by the teleportation probability in the PageRank.
Then, why do we feel the need to introduce a new centrality?

First, the extended influence inherits 
the definition of the influence for strongly connected networks 
and one-root networks (i.e., influence when $q=0$), and therefore, it 
represents the importance of nodes
in various dynamics and in the enumeration of spanning trees
(\SEC\ref{sec:interpretation}). 
Actually, for each dynamics considered in 
\SEC\ref{sec:interpretation}, we can consider a discrete-time version
and relate the importance of nodes in the dynamics to the PageRank.
We have explained this correspondence for the random walk
(\SEC\ref{sec:our def}). In addition,
the DeGroot model of opinion formation was originally proposed
in discrete time \cite{Degroot74-Friedkin91}. We should
choose one among the two centrality
measures depending on whether 
the continuous-time or
discrete-time dynamics are assumed to occur on
the network in question. 

In the discrete-time interpretation, the indegree is essentially
normalized to be unity. Therefore, if the weight of the link
represents a value that should not be normalized, such as the rate of
interaction, nominal connection strength, amount of signal or monetary
fluxes, and the number of wins and losses between a pair of sports
teams, the continuous-time interpretation, that is, the influence,
appears to be more appropriate. On the other hand,
the PageRank is more appropriate in
the case of scientometry; if
a paper cites many papers, the value of each citation should
be considered to be small, and
being cited from this paper should not be of great importance.
This distinction may underlie the current situation that
the PageRank and the Laplacian-based centrality
have been used in somewhat different research communities and for
different types of data. In this light, we have extended the
Laplacian-based centrality so that it is applicable to
general directed
networks, as is the PageRank.

Second, the PageRank has a subtle arbitrariness in determining the
behavior of the random walk that has reached a dangling node.
Depending on the implementation, the walker at a dangling node hops to a
randomly chosen node even with probability $1-q$ ($0<q\le 1$)
\cite{Brin98-Langville06book} or stays at the same node with probability
$1-q$ \cite{Fortunato06pnas}. 
The theoretical justification for either assumption
is not clear.  In the influence, we have the sole control
parameter $q$, and  the influence unambiguously 
corresponds to the discrete-time case in which the walker stays at the
dangling node with probability $1-q$.

\begin{acknowledgments}
We thank Yoji Kawamura for the helpful discussions and for
careful reading of the paper.
N.M. acknowledges the support provided by
the Grants-in-Aid for Scientific Research
(Grant Nos. 20760258 and 20540382, and
Innovative Areas ``Systems Molecular Ethology'') 
from MEXT, Japan.
\end{acknowledgments}

\section*{Appendix: 
Influence for two toy networks with multiple root nodes}

\begin{figure}
\begin{center}
\includegraphics[width=8cm]{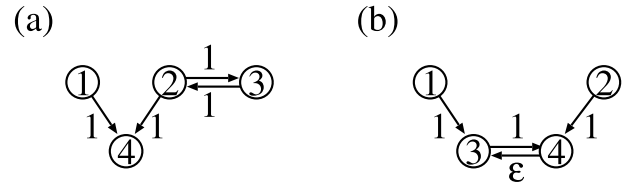}
\caption{Two directed networks with two root nodes.}
\label{fig:toy examples}
\end{center}
\end{figure}

For the network with four nodes and two root nodes
shown in \FIG\ref{fig:toy examples}(a), we obtain
$m=2$, $b=3$, $N_1=1$, $N_2=2$, $N_3=1$,
\begin{equation}
L=\left(\begin{array}{cccc}
0&0&0&0\\
0&1&-1&0\\
0&-1&1&0\\
-1&-1&0&2
\end{array}\right),
\end{equation}
\begin{alignat}{2}
\bm v^{(0)}_1 &= \left(1\; 0\; 0\; 0\right),&\quad 
\bm u^{(0)}_1 &= \left(\begin{array}{c}
1\\ 0\\ 0\\ \frac{1}{2}\end{array}\right),\\
\bm v^{(0)}_2 &= \left(0\; \frac{1}{2}\; \frac{1}{2}\; 0\right),&\quad 
\bm u^{(0)}_2 &= \left(\begin{array}{c}
0\\ 1\\ 1\\ \frac{1}{2}\end{array}\right).
\end{alignat}
Therefore, the influence is given by
\begin{equation}
v_1=\frac{3}{8},\quad v_2=v_3=\frac{5}{16},\quad v_4=0.
\end{equation}
Nodes 2 and 3 have the same influence because they are as strong as
each other within their SCC. Although the two upstream SCCs are upstream to
node 4 in the same manner, $v_1$ is smaller than $v_2+v_3$ because
$v_1$ controls two nodes and the SCC of nodes
2 and 3 controls three nodes.
 
For the network with four nodes and two root nodes
shown in \FIG\ref{fig:toy examples}(b), we obtain
$m=2$, $b=3$, $N_1=1$, $N_2=1$, $N_3=2$,
\begin{equation}
L=\left(\begin{array}{cccc}
0&0&0&0\\
0&0&0&0\\
-1&0&1+\epsilon&-\epsilon\\
0&-1&-1&2
\end{array}\right),
\end{equation}
\begin{alignat}{2}
\bm v^{(0)}_1 &= (1\; 0\; 0\; 0),&\quad
\bm u^{(0)}_1 &= \left(\begin{array}{c}
1\\ 0\\ \frac{2}{2+\epsilon}\\
\frac{1}{2+\epsilon}\end{array}\right),\\
\bm v^{(0)}_2 &= (0\; 1\; 0\; 0),&\quad
\bm u^{(0)}_2 &= \left(\begin{array}{c}
0\\ 1\\ \frac{\epsilon}{2+\epsilon}\\
\frac{1+\epsilon}{2+\epsilon}\end{array}\right).
\end{alignat}
The influence is given by
\begin{equation}
v_1=\frac{5+\epsilon}{4(2+\epsilon)},\quad 
v_2=\frac{3(1+\epsilon)}{4(2+\epsilon)},\quad v_3=v_4=0.
\end{equation}
$v_1>v_2$ because node 1 is connected to the more influential node
of the downstream SCC (i.e., $v_3$) unlike
node 2, which links to the less influential node of the downstream
SCC (i.e., $v_4$).
Note that when $\epsilon\approx 0$, the effect of node 1 on node 4
is similar to that of node 2 on node 4, despite the fact that
node 1 does not directly link to node 4, whereas node 2 does.
The reverse is not the case; when $\epsilon\approx 0$,
node 1 can affect node 3, but node 2 can hardly do so.

\newpage
\clearpage

\clearpage

\clearpage

\clearpage

\clearpage

\clearpage

\clearpage

\begin{thebibliography}{40}
\expandafter\ifx\csname natexlab\endcsname\relax\def\natexlab#1{#1}\fi
\expandafter\ifx\csname bibnamefont\endcsname\relax
  \def\bibnamefont#1{#1}\fi
\expandafter\ifx\csname bibfnamefont\endcsname\relax
  \def\bibfnamefont#1{#1}\fi
\expandafter\ifx\csname citenamefont\endcsname\relax
  \def\citenamefont#1{#1}\fi
\expandafter\ifx\csname url\endcsname\relax
  \def\url#1{\texttt{#1}}\fi
\expandafter\ifx\csname urlprefix\endcsname\relax\def\urlprefix{URL }\fi
\providecommand{\bibinfo}[2]{#2}
\providecommand{\eprint}[2][]{\url{#2}}

\bibitem[{\citenamefont{Albert and Barab\'{a}si}(2002)}]{albert02rmp}
\bibinfo{author}{\bibfnamefont{R.}~\bibnamefont{Albert}} \bibnamefont{and}
  \bibinfo{author}{\bibfnamefont{A.-L.} \bibnamefont{Barab\'{a}si}},
  \bibinfo{journal}{Rev. Mod. Phys.} \textbf{\bibinfo{volume}{74}},
  \bibinfo{pages}{47} (\bibinfo{year}{2002}).

\bibitem[{\citenamefont{Boccaletti et~al.}(2006)\citenamefont{Boccaletti,
  Latora, Moreno, Chavez, and Hwang}}]{Boccaletti06}
\bibinfo{author}{\bibfnamefont{S.}~\bibnamefont{Boccaletti}},
  \bibinfo{author}{\bibfnamefont{V.}~\bibnamefont{Latora}},
  \bibinfo{author}{\bibfnamefont{Y.}~\bibnamefont{Moreno}},
  \bibinfo{author}{\bibfnamefont{M.}~\bibnamefont{Chavez}}, \bibnamefont{and}
  \bibinfo{author}{\bibfnamefont{D.-U.} \bibnamefont{Hwang}},
  \bibinfo{journal}{Phys. Rep.} \textbf{\bibinfo{volume}{424}},
  \bibinfo{pages}{175} (\bibinfo{year}{2006}).

\bibitem[{\citenamefont{Newman}(2010)}]{Newman10book}
\bibinfo{author}{\bibfnamefont{M.~E.~J.} \bibnamefont{Newman}},
  \emph{\bibinfo{title}{Networks --- An introduction}}
  (\bibinfo{publisher}{Oxford University Press}, \bibinfo{address}{Oxford},
  \bibinfo{year}{2010}).

\bibitem[{\citenamefont{Katz}(1953)}]{Katz53}
\bibinfo{author}{\bibfnamefont{L.}~\bibnamefont{Katz}},
  \bibinfo{journal}{Psychometrika} \textbf{\bibinfo{volume}{18}},
  \bibinfo{pages}{39} (\bibinfo{year}{1953}).

\bibitem{Wasserman94-Brandes05LNCSbook}
\bibinfo{author}{\bibfnamefont{S.}~\bibnamefont{Wasserman}} \bibnamefont{and}
  \bibinfo{author}{\bibfnamefont{K.}~\bibnamefont{Faust}},
  \emph{\bibinfo{title}{Social Network Analysis}}
  (\bibinfo{publisher}{Cambridge University Press}, \bibinfo{address}{New
  York}, \bibinfo{year}{1994});
%
\emph{\bibinfo{title}{Network Analysis}}, edited by
\bibinfo{author}{\bibfnamefont{U.}~\bibnamefont{Brandes}} \bibnamefont{and}
  \bibinfo{author}{\bibfnamefont{T.~Erlebach}}
   (\bibinfo{publisher}{Springer},
  \bibinfo{address}{Berlin}, \bibinfo{year}{2005}).

\bibitem{Brin98-Langville06book}
\bibinfo{author}{\bibfnamefont{S.}~\bibnamefont{Brin}} \bibnamefont{and}
  \bibinfo{author}{\bibfnamefont{L.}~\bibnamefont{Page}},
  \bibinfo{journal}{Proceedings of the Seventh International World Wide Web
  Conference} pp. \bibinfo{pages}{107--117} (Elsevier, Amsterdam, \bibinfo{year}{1998});
\bibinfo{author}{\bibfnamefont{A.~N.} \bibnamefont{Langville}}
  \bibnamefont{and} \bibinfo{author}{\bibfnamefont{C.~D.} \bibnamefont{Meyer}},
  \emph{\bibinfo{title}{Google's PageRank and Beyond}}
  (\bibinfo{publisher}{Princeton University Press},
  \bibinfo{address}{Princeton}, \bibinfo{year}{2006}).

\bibitem{Daniels69Biom-Berman80SIAM}
\bibinfo{author}{\bibfnamefont{H.~E.} \bibnamefont{Daniels}},
  \bibinfo{journal}{Biometrika} \textbf{\bibinfo{volume}{56}},
  \bibinfo{pages}{295} (\bibinfo{year}{1969});
\bibinfo{author}{\bibfnamefont{K.~A.} \bibnamefont{Berman}},
  \bibinfo{journal}{SIAM J. Algebraic Disccrete Methods} \textbf{\bibinfo{volume}{1}},
  \bibinfo{pages}{359} (\bibinfo{year}{1980}).

\bibitem[{\citenamefont{Moon and Pullman}(1970)}]{Moon70SiamRev}
\bibinfo{author}{\bibfnamefont{J.~W.} \bibnamefont{Moon}} \bibnamefont{and}
  \bibinfo{author}{\bibfnamefont{N.~J.} \bibnamefont{Pullman}},
  \bibinfo{journal}{SIAM Rev.} \textbf{\bibinfo{volume}{12}},
  \bibinfo{pages}{384} (\bibinfo{year}{1970}).

\bibitem[{\citenamefont{Biggs}(1997)}]{Biggs97BullLondMathSoc}
\bibinfo{author}{\bibfnamefont{N.}~\bibnamefont{Biggs}},
  \bibinfo{journal}{Bull. London Math. Soc.} \textbf{\bibinfo{volume}{29}},
  \bibinfo{pages}{641} (\bibinfo{year}{1997}).

\bibitem[{\citenamefont{Agaev and Chebotarev}(2000)}]{Agaev00}
\bibinfo{author}{\bibfnamefont{R.~P.} \bibnamefont{Agaev}} \bibnamefont{and}
  \bibinfo{author}{\bibfnamefont{P.~Y.} \bibnamefont{Chebotarev}},
  \bibinfo{journal}{Autom. Remote Control (Engl. Transl.)} \textbf{\bibinfo{volume}{61}},
  \bibinfo{pages}{1424} (\bibinfo{year}{2000}).

\bibitem[{\citenamefont{Chebotarev and Agaev}(2005)}]{Chebotarev05arxiv}
\bibinfo{author}{\bibfnamefont{P.}~\bibnamefont{Chebotarev}} \bibnamefont{and}
  \bibinfo{author}{\bibfnamefont{R.}~\bibnamefont{Agaev}},
  \bibinfo{pages}{e-print arXiv:math/0508171}

\bibitem[{\citenamefont{Borm}(2002)}]{Borm02}
\bibinfo{author}{\bibfnamefont{N.~E.} \bibnamefont{Borm}},
\bibinfo{author}{\bibfnamefont{R.~Van Den Brink}},
 \bibnamefont{and}
  \bibinfo{author}{\bibfnamefont{M.}~\bibnamefont{Slikker}},
  \bibinfo{journal}{Ann. Operat. Res.} \textbf{\bibinfo{volume}{109}},
  \bibinfo{pages}{61} (\bibinfo{year}{2002}).

\bibitem[{\citenamefont{Masuda et~al.}(2009{\natexlab{a}})\citenamefont{Masuda,
  Kawamura, and Kori}}]{mkk09njp}
\bibinfo{author}{\bibfnamefont{N.}~\bibnamefont{Masuda}},
  \bibinfo{author}{\bibfnamefont{Y.}~\bibnamefont{Kawamura}}, \bibnamefont{and}
  \bibinfo{author}{\bibfnamefont{H.}~\bibnamefont{Kori}}, \bibinfo{journal}{New
  J. Phys.} \textbf{\bibinfo{volume}{11}}, \bibinfo{pages}{113002}
  (\bibinfo{year}{2009}{\natexlab{a}}).

\bibitem[{\citenamefont{Masuda et~al.}(2009{\natexlab{b}})\citenamefont{Masuda,
  Kawamura, and Kori}}]{mkk09pre}
\bibinfo{author}{\bibfnamefont{N.}~\bibnamefont{Masuda}},
  \bibinfo{author}{\bibfnamefont{Y.}~\bibnamefont{Kawamura}}, \bibnamefont{and}
  \bibinfo{author}{\bibfnamefont{H.}~\bibnamefont{Kori}},
  \bibinfo{journal}{Phys. Rev. E} \textbf{\bibinfo{volume}{80}},
  \bibinfo{pages}{046114} (\bibinfo{year}{2009}{\natexlab{b}}).

\bibitem[{\citenamefont{Klemm et~al.}(2010)\citenamefont{Klemm, Serrano,
  Egu\'{\i}luz, and Miguel}}]{Klemm10arxiv}
\bibinfo{author}{\bibfnamefont{K.}~\bibnamefont{Klemm}},
  \bibinfo{author}{\bibfnamefont{M.~\'{A}.}~\bibnamefont{Serrano}},
  \bibinfo{author}{\bibfnamefont{V.~M.} \bibnamefont{Egu\'{\i}luz}},
  \bibnamefont{and} \bibinfo{author}{\bibfnamefont{M.~San~Miguel}}, \bibinfo{pages}{arXiv:1002.4042v2}

\bibitem[{\citenamefont{Golub and Loan}(1996)}]{Golub96book}
\bibinfo{author}{\bibfnamefont{G.~H.} \bibnamefont{Golub}} \bibnamefont{and}
  \bibinfo{author}{\bibfnamefont{C.~F.~V.} \bibnamefont{Loan}},
  \emph{\bibinfo{title}{Matrix Computations, 3rd ed.}}
  (\bibinfo{publisher}{Johns Hopkins University Press},
  \bibinfo{address}{Baltimore}, \bibinfo{year}{1996}).

\bibitem[{\citenamefont{Horn and Johnson}(1985)}]{Hornbook}
\bibinfo{author}{\bibfnamefont{R.~A.} \bibnamefont{Horn}} \bibnamefont{and}
  \bibinfo{author}{\bibfnamefont{C.~R.} \bibnamefont{Johnson}},
  \emph{\bibinfo{title}{Matrix Analysis}} (\bibinfo{publisher}{Cambridge University
  Press}, \bibinfo{address}{Cambridge, England}, \bibinfo{year}{1985}).

\bibitem[{\citenamefont{Abramson}(1964)}]{Abelson64chap}
\bibinfo{author}{\bibfnamefont{R.~P.} \bibnamefont{Abramson}},
  \bibinfo{journal}{in Contributions to Mathematical Psychology}, edited by N.
  Frederiksen and H. Gulliksen, (Holt, Rinehart and Winston, New York, 1964), pp. \bibinfo{pages}{141--160}.

\bibitem[{\citenamefont{Ermentrout}(1992)}]{ermentrout92}
\bibinfo{author}{\bibfnamefont{G.~B.}~\bibnamefont{Ermentrout}},
  \bibinfo{journal}{SIAM J. Appl. Math.} \textbf{\bibinfo{volume}{52}},
  \bibinfo{pages}{1665} (\bibinfo{year}{1992}).

\bibitem[{\citenamefont{Masuda and Ohtsuki}(2009)}]{MasudaOhtsuki09njp}
\bibinfo{author}{\bibfnamefont{N.}~\bibnamefont{Masuda}} \bibnamefont{and}
  \bibinfo{author}{\bibfnamefont{H.}~\bibnamefont{Ohtsuki}},
  \bibinfo{journal}{New J. Phys.} \textbf{\bibinfo{volume}{11}},
  \bibinfo{pages}{033012} (\bibinfo{year}{2009}).

\bibitem[{\citenamefont{Bapat and Raghavan}(1997)}]{Bapat97book}
\bibinfo{author}{\bibfnamefont{R.~B.} \bibnamefont{Bapat}} \bibnamefont{and}
  \bibinfo{author}{\bibfnamefont{T.~E.~S.} \bibnamefont{Raghavan}},
  \emph{\bibinfo{title}{Nonnegative Matrices and Applications}}
  (\bibinfo{publisher}{Cambridge University Press},
  \bibinfo{address}{Cambridge, England}, \bibinfo{year}{1997}).

\bibitem[{\citenamefont{Berman and Plemmons}(1979)}]{BermanPlemmons79book}
\bibinfo{author}{\bibfnamefont{A.}~\bibnamefont{Berman}} \bibnamefont{and}
  \bibinfo{author}{\bibfnamefont{R.~J.} \bibnamefont{Plemmons}},
  \emph{\bibinfo{title}{Nonnegative Matrices in the Mathematical Sciences}}
  (\bibinfo{publisher}{Academic Press}, \bibinfo{address}{San Diego, CA},
  \bibinfo{year}{1979}).

\bibitem{Degroot74-Friedkin91}
\bibinfo{author}{\bibfnamefont{M.~H.} \bibnamefont{DeGroot}},
  \bibinfo{journal}{J. Am. Stat. Assoc.} \textbf{\bibinfo{volume}{69}},
  \bibinfo{pages}{118} (\bibinfo{year}{1974});
\bibinfo{author}{\bibfnamefont{N.~E.} \bibnamefont{Friedkin}},
  \bibinfo{journal}{Am. J. Sociol.} \textbf{\bibinfo{volume}{96}},
  \bibinfo{pages}{1478} (\bibinfo{year}{1991}).

\bibitem[{\citenamefont{Olfati-Saber et~al.}(2007)\citenamefont{Olfati-Saber,
  Fax, and Murray}}]{olfatisaber07}
\bibinfo{author}{\bibfnamefont{R.}~\bibnamefont{Olfati-Saber}},
  \bibinfo{author}{\bibfnamefont{J.}~\bibnamefont{Fax}}, \bibnamefont{and}
  \bibinfo{author}{\bibfnamefont{R.}~\bibnamefont{Murray}},
  \bibinfo{journal}{Proc. IEEE} \textbf{\bibinfo{volume}{95}},
  \bibinfo{pages}{215} (\bibinfo{year}{2007}).

\bibitem[{\citenamefont{Kori et~al.}(2009)\citenamefont{Kori, Kawamura, Nakao,
  Arai, and Kuramoto}}]{kori09}
\bibinfo{author}{\bibfnamefont{H.}~\bibnamefont{Kori}},
  \bibinfo{author}{\bibfnamefont{Y.}~\bibnamefont{Kawamura}},
  \bibinfo{author}{\bibfnamefont{H.}~\bibnamefont{Nakao}},
  \bibinfo{author}{\bibfnamefont{K.}~\bibnamefont{Arai}}, \bibnamefont{and}
  \bibinfo{author}{\bibfnamefont{Y.}~\bibnamefont{Kuramoto}},
  \bibinfo{journal}{Phys. Rev. E} \textbf{\bibinfo{volume}{80}},
  \bibinfo{pages}{036207} (\bibinfo{year}{2009}).

\bibitem{Antal06prl-Sood08pre}
\bibinfo{author}{\bibfnamefont{T.}~\bibnamefont{Antal}},
  \bibinfo{author}{\bibfnamefont{S.}~\bibnamefont{Redner}}, \bibnamefont{and}
  \bibinfo{author}{\bibfnamefont{V.}~\bibnamefont{Sood}},
  \bibinfo{journal}{Phys. Rev. Lett.} \textbf{\bibinfo{volume}{96}},
  \bibinfo{pages}{188104} (\bibinfo{year}{2006});
\bibinfo{author}{\bibfnamefont{V.}~\bibnamefont{Sood}},
  \bibinfo{author}{\bibfnamefont{T.}~\bibnamefont{Antal}}, \bibnamefont{and}
  \bibinfo{author}{\bibfnamefont{S.}~\bibnamefont{Redner}},
  \bibinfo{journal}{Phys. Rev. E} \textbf{\bibinfo{volume}{77}},
  \bibinfo{pages}{041121} (\bibinfo{year}{2008}).

\bibitem[{\citenamefont{Leighton and Rivest}(1986)}]{Leighton86IEEE}
\bibinfo{author}{\bibfnamefont{F.~T.} \bibnamefont{Leighton}} \bibnamefont{and}
  \bibinfo{author}{\bibfnamefont{R.~L.} \bibnamefont{Rivest}},
  \bibinfo{journal}{IEEE Trans. Inf. Theory} \textbf{\bibinfo{volume}{32}},
  \bibinfo{pages}{733} (\bibinfo{year}{1986}).

\bibitem{LiuGao08ACM-LiuLiu10InfRetrieval}
\bibinfo{author}{\bibfnamefont{Y.}~\bibnamefont{Liu}},
  \bibinfo{author}{\bibfnamefont{B.}~\bibnamefont{Gao}},
  \bibinfo{author}{\bibfnamefont{T.~Y.} \bibnamefont{Liu}},
  \bibinfo{author}{\bibfnamefont{Y.}~\bibnamefont{Zhang}},
  \bibinfo{author}{\bibfnamefont{Z.}~\bibnamefont{Ma}},
  \bibinfo{author}{\bibfnamefont{S.}~\bibnamefont{He}}, \bibnamefont{and}
  \bibinfo{author}{\bibfnamefont{H.}~\bibnamefont{Li}}, \bibinfo{journal}{
  SIGIR’08: Proceedings of the 31st Annual International ACM SIGIR Conference
  on Research and Development in Information Retrieval (ACM, New York, 2008)} pp.
  \bibinfo{pages}{451--458};
\bibinfo{author}{\bibfnamefont{Y.}~\bibnamefont{Liu}},
  \bibinfo{author}{\bibfnamefont{T.~Y.} \bibnamefont{Liu}},
  \bibinfo{author}{\bibfnamefont{B.}~\bibnamefont{Gao}},
  \bibinfo{author}{\bibfnamefont{Z.}~\bibnamefont{Ma}}, \bibnamefont{and}
  \bibinfo{author}{\bibfnamefont{H.}~\bibnamefont{Li}}, \bibinfo{journal}{Inf.
  Retrieval} \textbf{\bibinfo{volume}{13}}, \bibinfo{pages}{22}
  (\bibinfo{year}{2010}).

\bibitem{Avrachenkov2007siam}
K. Avrachenkov, N. Litvak, D. Nemirovsky, and N. Osipova,
SIAM J. Numer. Anal. \textbf{45}, 890 (2007).

\bibitem{Chen06pnas-wormatlas}
\bibinfo{author}{\bibfnamefont{B.~L.} \bibnamefont{Chen}},
  \bibinfo{author}{\bibfnamefont{D.~H.} \bibnamefont{Hall}}, \bibnamefont{and}
  \bibinfo{author}{\bibfnamefont{D.~B.} \bibnamefont{Chklovskii}},
  \bibinfo{journal}{Proc. Natl. Acad. Sci. U.S.A.} \textbf{\bibinfo{volume}{103}},
  \bibinfo{pages}{4723} (\bibinfo{year}{2006});
\emph{\bibinfo{title}{{\rm http://www.wormatlas.org}}}.

\bibitem[{\citenamefont{Panzarasa et~al.}(2009)\citenamefont{Panzarasa, Opsahl,
  and Carley}}]{Panzarasa09JASIST}
\bibinfo{author}{\bibfnamefont{P.}~\bibnamefont{Panzarasa}},
  \bibinfo{author}{\bibfnamefont{T.}~\bibnamefont{Opsahl}}, \bibnamefont{and}
  \bibinfo{author}{\bibfnamefont{K.~M.} \bibnamefont{Carley}},
  \bibinfo{journal}{J. Am. Soc. Inf. Sci. Technol.}
  \textbf{\bibinfo{volume}{60}}, \bibinfo{pages}{911} (\bibinfo{year}{2009}).

\bibitem{Palacios04-Fersht09}
\bibinfo{author}{\bibfnamefont{I.}~\bibnamefont{Palacios-Huerta}}
  \bibnamefont{and} \bibinfo{author}{\bibfnamefont{O.}~\bibnamefont{Volij}},
  \bibinfo{journal}{Econometrica} \textbf{\bibinfo{volume}{72}},
  \bibinfo{pages}{963} (\bibinfo{year}{2004});
\bibinfo{author}{\bibfnamefont{A.}~\bibnamefont{Fersht}},
  \bibinfo{journal}{Proc. Natl. Acad. Sci. U.S.A.} \textbf{\bibinfo{volume}{106}},
  \bibinfo{pages}{6883} (\bibinfo{year}{2009}).

\bibitem[{\citenamefont{Fortunato et~al.}(2006)\citenamefont{Fortunato,
  Flammini, Menczer, and Vespignani}}]{Fortunato06pnas}
\bibinfo{author}{\bibfnamefont{S.}~\bibnamefont{Fortunato}},
  \bibinfo{author}{\bibfnamefont{A.}~\bibnamefont{Flammini}},
  \bibinfo{author}{\bibfnamefont{F.}~\bibnamefont{Menczer}}, \bibnamefont{and}
  \bibinfo{author}{\bibfnamefont{A.}~\bibnamefont{Vespignani}},
  \bibinfo{journal}{Proc. Natl. Acad. Sci. U.S.A.} \textbf{\bibinfo{volume}{103}},
  \bibinfo{pages}{12684} (\bibinfo{year}{2006}).

\end{thebibliography}
\end{document}